\documentclass{aa}  
\usepackage{graphicx}
\usepackage{txfonts}
\usepackage{multirow}
\usepackage{color}
\usepackage{tabularx}

\usepackage{ulem}

\def\Box{\mathord{\dalemb{7.9}{8}\hbox{\hskip1pt}}}
\def\dalemb#1#2{{\vbox{\hrule height.#2pt
  \hbox{\vrule width.#2pt height#1pt \kern#1pt \vrule width.#2pt}
    \hrule height.#2pt}}}

\def\ba{\begin{eqnarray}}
\def\ea{\end{eqnarray}}
\def\be{\begin{equation}}
\def\ee{\end{equation}}

\def\gtorder{\mathrel{\raise.3ex\hbox{$>$}\mkern-14mu
             \lower0.6ex\hbox{$\sim$}}}
\def\ltorder{\mathrel{\raise.3ex\hbox{$<$}\mkern-14mu
             \lower0.6ex\hbox{$\sim$}}}

\begin{document}

%%%%%%%%%%%%%%%%%%%%%%%%%%%%%%%%%%%%%%%%%%%%%%%%%%

%%%%%%%%%%%%%%%%%%% TITLE PAGE %%%%%%%%%%%%%%%%%%%

   \title{
   Angular distribution of cosmological parameters as a probe of inhomogeneities: a kinematic parametrisation}
   %\subtitle{I. Overviewing the $\kappa$-mechanism}
   \author{C. Sofia Carvalho
          \inst{1,2}\fnmsep\thanks{Corresponding author: cscarvalho@oal.ul.pt}
          \and
          Spyros Basilakos\inst{2}
          }
   \institute{Institute of Astrophysics and Space Sciences, University of Lisbon, 
   Tapada da Ajuda, 1349--018 Lisbon, Portugal,
              %\\
              %\email{wuchterl@amok.ast.univie.ac.at}
         \and
             Research Center for Astronomy and Applied Mathematics, Academy of Athens, 
             Soranou Efessiou 4, 11--527, Athens, Greece %\\
             %\email{c.ptolemy@hipparch.uheaven.space}
             %\thanks{T$\Theta\Delta$}
             }
   %\date{Received September 15, 1996; accepted March 16, 1997}

\abstract{
We use a kinematic parametrisation of the luminosity distance to measure the angular distribution on the sky of
time derivatives of the scale factor, %namely
in particular
the Hubble parameter $H_0,$ the deceleration parameter $q_0,$ and the jerk parameter $j_0.$ We apply a recently published method to complement probing the inhomogeneity of the large--scale structure by means of the inhomogeneity in the cosmic expansion.
This parametrisation is independent of the cosmological equation of state,  which renders it adequate to test interpretations of the cosmic acceleration alternative to the cosmological constant.  
For the same analytical toy model of an inhomogeneous ensemble of homogenous pixels, we derive the backreaction term in $j_0$ due to the fluctuations of $\{H_0,q_0\}$ and measure it to be of order 
$10^{-2}$ 
times the corresponding average over the pixels in the absence of backreaction. 
In agreement with that computed using a $\Lambda$CDM parametrisation of the luminosity distance, the backreaction effect on $q_0$ remains below the detection threshold.
Although the backreaction effect on $j_0$ is about ten times that on $q_0,$ it is also below the detection threshold. 
Hence backreaction remains unobservable both in $q_0$ and in $j_0.$}

%\titlerunning{Angular distribution of cosmological parameters: a kinematic parametrisation}
\titlerunning{Angular distribution of cosmological parameters}
\authorrunning{Carvalho \& Basilakos}
\maketitle

%%%%%%%%%%%%%%%%%%%%%%%%%%%%%%%%%%%%%%%%%%%%%%%%%%

%%%%%%%%%%%%%%%%%%% BODY OF PAPER %%%%%%%%%%%%%%%%%%%

\section{Introduction}

Supernova (SN) data have provided evidence that distant sources ($z > 0.3$) appear dimmer than predicted in a universe with matter only, in comparison with nearby sources. This dimming led to the interpretation that the Universe is expanding in an accelerating fashion. This acceleration has been attributed to a dark energy component, whose simplest solution is a vacuum energy or equivalently a cosmological constant.
However, this dimming could be caused by a variation of any other component that affects the luminosity distance, namely by inhomogeneities in the energy densities or %by anisotropies 
in the cosmic expansion \citep{amendola_2013}.

In order to explain the underlying mechanism of the 
late--time accelerated expansion of the universe, one possibility is  
to give up of the cosmological principle and allow for an anisotropic 
expansion of the universe.
The idea that we live in a locally underdense region, hence creating a `Hubble 
bubble', can explain the cosmic acceleration at late times, subject to specific conditions \citep{zehavi_1998,caldwell_2008}.
From the vast literature on the cosmological principle, it has been found that most cosmological observations 
can accommodate violations of the cosmological principle 
(see for example \citet{zibin_2008,komita_2009,marra_2010,marra_2011,moss_2011}). 
From the theoretical point of view, violations of the cosmological principle
can be explained in the context of Lema\^itre--Tolman--Bondi (LTB) void models. These void models are spherically symmetric and radially inhomogeneous, and can 
%resemble the concordance $\Lambda$CDM as far as the cosmic expansion is concerned (namely $H(z),$ luminosity distance, angular distance, etc).
mimic the cosmic expansion of the concordance $\Lambda$CDM \citep{lan_2010, liu_2014}.

In \cite{carvalho_2015}, we introduced a method to probe the inhomogeneity of the large--scale structure %in the form of a local parameter estimation of cosmological parameters 
by measuring the angular distribution in the cosmological parameters that affect the luminosity distance, using SN data. Variation in the cosmological parameters across pixels in the sky implies inhomogeneity in the cosmic expansion. This inhomogeneity was then used to measure the extra component of cosmic acceleration predicted by backreaction, which derives from averaging over an inhomogeneous ensemble of homogeneous pixels, each pixel expanding at a different rate. %and hence to constrain an alternative mechanism to generate cosmic acceleration.
However, this measurement presupposed an a priori dark energy component in each pixel. In order to investigate alternative interpretations of the cosmic acceleration, it is conceptually more consistent to use a parametrisation that does not assume a specific component as the cause of acceleration \citep{riess_2002}.

%[Motivate kinematic record of expansion history of the universe without regard to cause]

In this manuscript, we use the luminosity distance expressed %in terms of cosmographic parameters, i.e. 
in terms of time derivatives of the scale factor $a(z),$ in particular the Hubble parameter $H_0,$ the deceleration parameter $q_0$ and the jerk parameter $j_0.$  
Instead of assuming an equation of state and inferring the evolution of the scale factor via the Friedmann equation, the reasoning is to take the data on the scale factor and  infer a cosmological equation of state via the Friedmann equation.
This parametrisation records the cosmic expansion without regard to its cause, thus being independent of the cosmological equation of state 
%and of the Friedmann equation describing $H(z)$
%and hence of a particular energy content of the Universe,
and consequently %renders this parametrisation 
adequate to test interpretations of the cosmic acceleration alternative to the cosmological constant.
These parameters can be related to the Taylor expansion of the cosmological equation of state about the present values $\{\rho_{0},p_{0}\}$ expressed up to linear order as 
\ba
p=p_{0}+\kappa_{0}(\rho-\rho_{0})+O[(\rho-\rho_{0})^2],
\ea
hence yielding information about the present values of $w_{0}$ and $\kappa_{0}$ defined as \citep{visser_2004}
\ba
w_{0}={p\over \rho}\Big \vert_{0}\equiv w_{0}(H_{0},q_{0}), \quad \kappa_{0}={dp\over d\rho}\Big \vert_{0}\equiv \kappa_{0}(H_{0},q_{0},j_{0}).
\ea
Whereas $w_{0}$ (and hence $q_0$) contains information about the present value of $p/\rho,$ $\kappa_{0}$ (and hence $j_0$) contains information about how $w_{0}$ can evolve.

The purpose of this manuscript is to reapply the method first presented in \citet{carvalho_2015} to estimate the parameters $\{H_0,q_0,j_0\},$ instead of $\{H_0,\Omega_{M},\Omega_{\Lambda}\},$ by fitting the luminosity distance to SN data, and to compare the results from the two estimations in view of an interpretation of the cosmic acceleration independent of a particular energy content of the Universe. (Although the comparisons are made for the case that $\Omega_{\kappa}=1-\Omega_{M}-\Omega_{\Lambda}$, 
for completion we also include the results for the case where $\Omega_{\kappa}=0.$) Some studies have used the kinematic parametrization to measure inhomogeneities from SN data (see e.g. \citet{schwarz_2007,kalus_2013}). Most studies, however, have aimed at finding hemispherical anisotropies assuming a $\Lambda$CDM energy content (e.g. \citet{blomqvist_2010, mariano_2012, heneka_2014, jimenez_2015, bengaly_2015, javanmardi_2015, migkas_2016} and references therein).

This paper is organised as follows. In Sec.~\ref{sec:param} we describe the data and estimate the observables by performing both a global and a local parameter estimation, obtaining fiducial values and maps  respectively for the estimated parameters. We introduce an inhomogeneity test by rotating the supernova subsampling per pixel. 
In Sec.~\ref{sec:power} we compute the power spectrum of the maps of the parameters using two methods for the noise bias removal.
In Sec.~\ref{sec:average}, for the same toy model of backreaction used in ~\citet{carvalho_2015}, we compute the average values of $\{q_0,j_0\}$ and discuss possible cosmological implications. In Sec.~\ref{sec:concl} we draw conclusions.

\section{Parameter estimation}
\label{sec:param}
%[Data]

We use the type Ia supernova sample compiled by the Joint Light--curve Analysis (JLA) collaborative effort \citep{betoule_2014} from different supernova surveys, totalling $N_{\text{SNe}}=740$ supernovae (SNe) with redshift 
$z\in [0.010,1.30]$ 
and distributed on the sky according to Fig.~1 in \citet{carvalho_2015}.
\footnote{The sample was obtained from {\tt http://supernovae.in2p3.fr/sdss\_snls\_jla/ReadMe.html}.} 

%[Estimate parameter in $d_L$ with $a(z)$ expressed as a Taylor expansion about $a(z_0)$] 

We use the luminosity distance $d_L$ with $a(z)$ expressed as a Taylor expansion about the present value $a(z_0)$ \citep{visser_2004}
\ba
&&d_{L}(z;H_0,q_0,j_0)={cz\over H_{0}}\times\cr
&\times&\!\!\!\!\Biggl[
1+{1\over 2}(1-q_0)z
-{1\over 6}\left(1-q_0-3q_0^2+j_0+{kc^2\over{H_0^2 a_0^2}}\right)z^2+O(z^3)\Biggr],\cr
&&
\label{eqn:dl}
\ea
to estimate the cosmological parameters that minimise the chi--square of the fit 
of the theoretical distance modulus
\ba
\mu_{\rm theo}(z;H_{0},q_0,j_0)
=5\log[d_{L}(z;H_{0},q_0,j_0)]+25
\label{eqn:mu_theo}
,\ea
($d_L$ in units of Mpc) computed for the trial values of $\{H_{0},q_{0}, j_{0}\},$ 
to the measured distance modulus
\ba
\mu_{\rm data}(z)=m_{B}(z)-(M_{B}-\alpha x_{1}+\beta c),
\label{eqn:mu_data}
\ea
computed for the light--curve parameters $\{m_{B},x_{1},c\}$ estimated from each SN's observed magnitude, and for $M_B=-19.05\pm 0.02,$ $\alpha=0.141\pm 0.006$ and $\beta=3.101\pm 0.075$ estimated for all SNe \citep{betoule_2014}. 

We recall that the $\{H_0,q_0\}$ expansion in Eq.~(\ref{eqn:dl}) (i.e. up to the second term in the right--hand side) does not allow for an estimation of $q_0$ that is both accurate and precise \citep{neben_2013}. This is because for low redshifts where the Taylor expansion is most accurate, there is poor leverage for a precise estimation of $q_0;$  conversely, for high redshifts where the leverage is larger and allows for a more precise estimation, the Taylor expansion is less accurate. Considering the $\{H_0,q_0,j_0\}$ expansion in Eq.~(\ref{eqn:dl}) (i.e. all the terms in the right--hand side), then we expect that the uncertainty will be pushed to the estimation of $j_0.$  
An expansion in higher--order time derivatives of the expansion factor was considered in \citet{aviles_2012}. An exhaustive comparison of luminosity distances can be found in \citet{cattoen_2008}. 

\begin{figure}
\vskip-1.5cm
\centerline{
\includegraphics[width=10.5cm]%{figspyros2.pdf}
{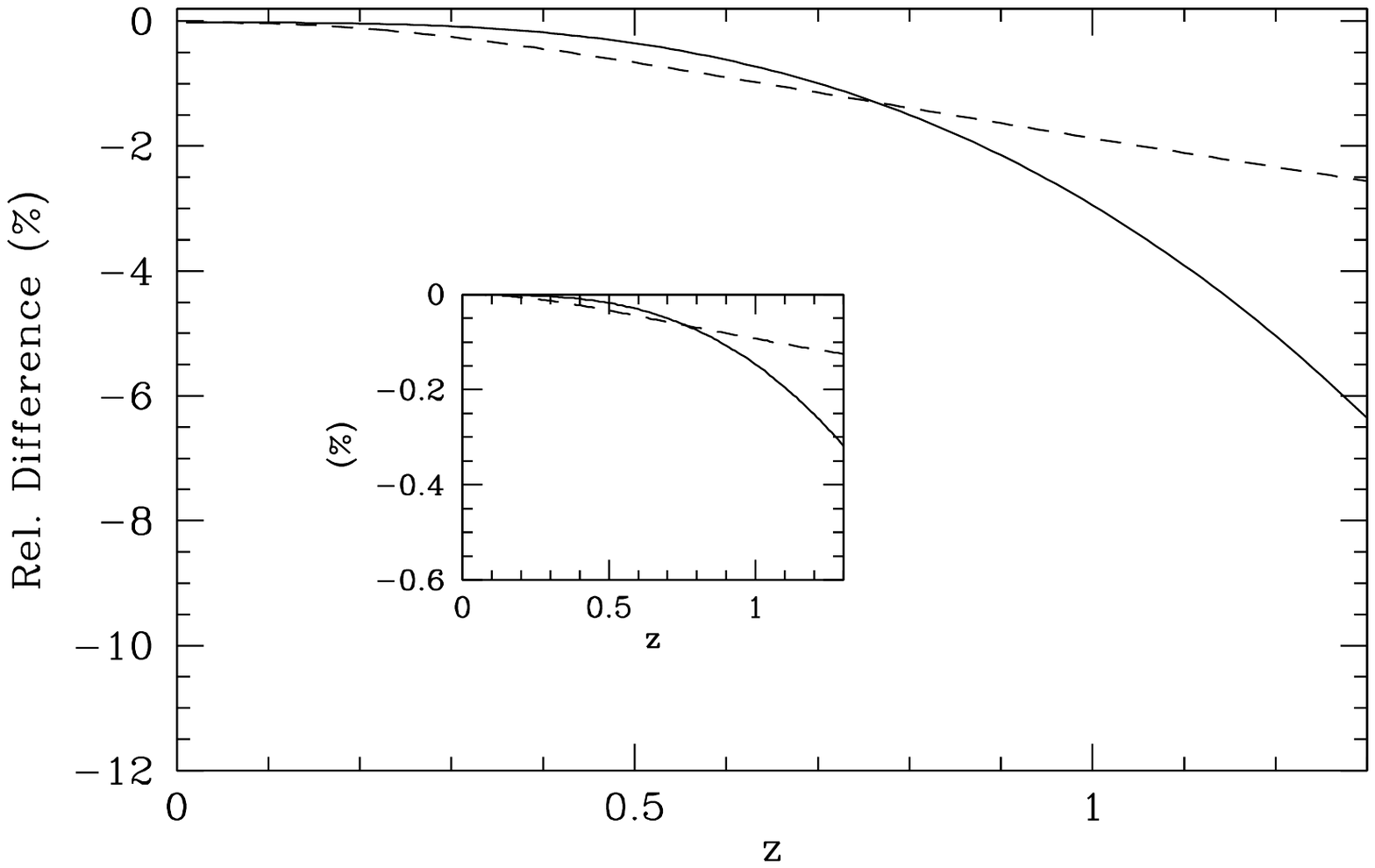}
}
\vskip-7.cm
\caption{{\bf Relative difference between expressions of the luminosity distance.}
The solid line shows the relative difference between the luminosity distance $d_{L}(z)$ in Eq.~(\ref{eqn:dl}) 
and the theoretical luminosity distance in the $\Lambda$CDM model in Eq.~(\ref{eqn:dl_z}) for $\Omega_{M}=0.308$ and $H_0=67.8~\text{km}~\text{s}^{-1}\text{Mpc}^{-1}$ \citep{planck_2015}. 
The dashed line shows the relative difference between the luminosity distance $d_L(y),$ with $y\equiv z/(1 + z),$ in \citet{cattoen_2008} and the theoretical luminosity distance in the $\Lambda$CDM model. The inset plot shows the corresponding differences in $\mu,$ given by Eq.~(\ref{eqn:mu_theo}), which are indicated by the corresponding line styles.}
\label{fig:spyros}
\end{figure}

Before proceeding further, we check the validity range of Eq.~(\ref{eqn:dl}) using a spatially 
flat $\Lambda$CDM model as reference model. In this case, we have 
$q_{0}=(\Omega_{M}/2)-\Omega_{\Lambda}$ and $j_{0}=\Omega_{M}+\Omega_{\Lambda}=1.$
We compute the relative difference 
between the luminosity distance $d_{L}(z)$ in Eq.~(\ref{eqn:dl}) and the theoretical luminosity distance 
in the $\Lambda$CDM model 
\ba
d_{L}(z)=\frac{c(1+z)}{H_{0}} \int_{0}^{z}\frac{dz}{E(z)},
\label{eqn:dl_z}
\ea
where $E(z)=[\Omega_{M}(1+z)^{3}+\Omega_{\Lambda}]^{1/2}$ for $\Omega_{M}=0.308$ and $H_0=67.8~\text{km}~\text{s}^{-1}\text{Mpc}^{-1}$ \citep{planck_2015}. (See solid line in Fig.~\ref{fig:spyros}.)
We find that, for $z\in [0.1,1.1],$ the relative difference lies in the interval $[-0.02\%,-3.9\%];$ conversely, for $z\in ]1.1,1.30],$ the relative difference can reach about $-6.4\%.$ %when $z\simeq 1.30.$ 
We also compute the relative difference between the luminosity distance $d_L(y),$ with $y\equiv z/(1 + z),$ in \citet{cattoen_2008} and the theoretical luminosity distance in the $\Lambda$CDM model. 
(See dashed line in Fig.~\ref{fig:spyros}.)
In this case we find that, for $z\in [0.1,1.1],$ the relative difference lies in the interval $[-0.02\%,-2.1\%];$ conversely, for $z\in ]1.1,1.30],$ the relative difference is about $-2.6\%$. We also observe that, in the intermediate redshift range $z\in [0.1,0.75]$ where most SNe were detected, Eq.~(\ref{eqn:dl}) performs slightly better than $d_L(y);$ conversely, $d_L(y)$ clearly performs better for $z>1.$ 
We also compute the corresponding relative differences in $\mu,$ which is the quantity that we use in the subsequent estimation and which depends on $d_{L}$ logarithmically according to Eq.~(\ref{eqn:mu_theo}). (See inset plot of Fig.~\ref{fig:spyros}.) We observe that the relative differences in $\mu$ are less than $-0.4\%,$ which implies that the differences in the luminosity density translate in a negligible effect on our results.

We now proceed to estimate the parameters $\{H_0,q_0,j_0\}$ using the method described in Appendix A of \citet{carvalho_2015}. 
We assume that at the present epoch $c/H_0a_0\ll 1,$ regardless of the value of $k,$ which means that by setting the prior $k=0$ and estimating $j_0$ we are actually estimating $j_0+kc^2/(H_0a_0)^2$ \citep{neben_2013}.
For a model consisting of an incoherent mixture of matter, where each component is described by an equation of state $p_i=w_i \rho_i,$ this assumption implies that $j_0\ge q_0(1+2q_0).$ We also assume that the measurements of the magnitude (and consequently of $\mu$) at different redshifts are independent,  %i.e. $\mu_{\text{data}}(z)$ and $\sigma_{\mu_{\text{data}}}$ are independent, 
which is equivalent to assuming that the parameters' distribution is Gaussian as a first approximation. %and consequently to using the diagonal elements only in the covariance matrix.%and that the noise is Gaussian as a first approximation. 
We compute the error 
$\sigma_{\mu_{\text{data}}}$ by assuming uncorrelated errors for the observables and by error propagating according to Eq.~(\ref{eqn:mu_data}).

%\begin{table}[t]
\begin{table*}%[t]
\caption{%\baselineskip=0.5cm{
{\bf Values for the parameters estimated from the JLA type Ia SN sample.} %Column 1: Parameters estimated either directly or indirectly from the fit. Column 2: Values estimated from the complete SN sample. Columns 3--7: Values estimated by other collaborations from SN data.%}
}
\label{table:param}
\begin{tabular}{c||c|ccccc}
\hline\hline
Parameter 
& \multicolumn{1}{c|} {Complete} %Sample} 
& Carvalho et al. %\citet{carvalho_2015} 
& Carvalho et al. %\citet{carvalho_2015} 
%& Riess et al. 
%& Riess et al. 
& Caldwell et al. %\citet{caldwell_2004}  
& Riess et al. %\citet{riess_2011} 
& Neben et al. \\ %\citet{neben_2013}\\ 
& {sample}& $\Omega_{\kappa}=1-\Omega_M-\Omega_{\Lambda}$ %free %\citep{carvalho_2015} 
%& {Sample}& $\Omega_{\kappa}$ free %\citep{carvalho_2015} 
& $\Omega_{\kappa}=0$ %\citep{carvalho_2015} 
%&``gold" \citep{riess_2004} 
%& ``gold+silver" \citep{riess_2004} 
& %\citep{caldwell_2004} 
&$j_0=1$%\citep{riess_2011} 
&%\citep{neben_2013} 
\\ \hline %\hline
$H_{0}$ 
& $71.06\pm 0.46$ 
& $71.17\pm 0.44$ 
& $71.21\pm 0.33$
%& & 
& 
&$73.8\pm2.4$
&\\
$q_0$ 
& $-0.540\pm 0.094$ 
& $-0.586\pm 0.123$ & $-0.599\pm 0.020$ 
%& $[-1.3,-0.2]$ 
%& $[-1.4,-0.3]$
& $[-1.1,-0.2]$ 
&$-0.55$ %\pm??$
&$-0.64\pm0.14$
\\
$j_0$ 
& $0.533\pm 0.503$ 
& $0.971\pm 0.139$ & $1.000 \pm0.026$
%& $[-0.3,5.9]$ 
%& $[-0.1,6.4]$ 
& $[-0.5,3.9]$  
&$1$
&$1.4\pm0.8$\\
\hline
\end{tabular}
\tablefoot{Column 1: Parameters estimated either directly or indirectly from the fit. Column 2: Values estimated from the complete SN sample. Columns 3--7: Values estimated by other collaborations from SN data.
}
\end{table*}
%\end{table}

\subsection{Global parameter estimation}

We first perform a parameter estimation using the complete SN sample, called the global estimation and corresponding to ${\tt npixel}=1$ pixel, from which we estimate the maximum likelihood values for the parameters $x_{i}=\{H_{0},q_{0},j_{0}\}.$ 
(Whenever unspecified, $H_{0}$ is measured in units of $\text{km}~\text{s}^{-1}\text{Mpc}^{-1}.$)

For the global estimation, we ran $n_j=10$
realizations of Markov chains of length $10^{4}.$ 
The starting point of each realization is randomly generated. By removing the first 20\% of entries in the chains to keep the burnt--in phase only, thinning down the remaining 80\% to a half by removing one of each consecutive entry in order to remove correlations within the chain, and finally averaging over the various chains,  we obtain the following results: $x_{i}^{\text{fid}}=\{H_0^{\text{fid}},q_0^{\text{fid}},j_0^{\text{fid}}\}=\{71.06,-0.540,0.533\}\pm \{0.46,0.094,0.503\}$ (see Table~\ref{table:param}). We will use these results as the fiducial values in the subsequent calculations.
The errors contain the dispersion in each chain and the dispersion among the averages of the different chains added in quadrature. 

The measurements of $\{H_0,q_{0},j_{0}\}$ are consistent with 
the measurements of $\{H_{0},q_{0}=\Omega_{M}/2-\Omega_{\Lambda},j_{0}=\Omega_{M}+\Omega_{\Lambda}\}$ obtained in \citet{carvalho_2015},  
which we include in Table~\ref{table:param} for convenience. 
The measurements of $\{H_0,q_{0}\}$ are equally precise, favouring $q_{0}<0$ at over $5\sigma.$ The measurement of $j_{0}$ is comparatively less precise, as also observed in \citet{neben_2013} in the context of the kinematic parametrisation, but nonetheless it favours $j_0>0$ at $1\sigma.$ These results imply that, in the interval $z \in [0.01,1.30]$ covered by the SN sample, the cosmic expansion accelerated and that previously it had decelerated, 
meaning that there was a time when the acceleration changed sign, which supports the evidence found in \citet{riess_2004}.

\begin{figure*}
\centerline{
\includegraphics[width=9cm]
%{jla_mu_vs_z_q0j0_col.pdf}
{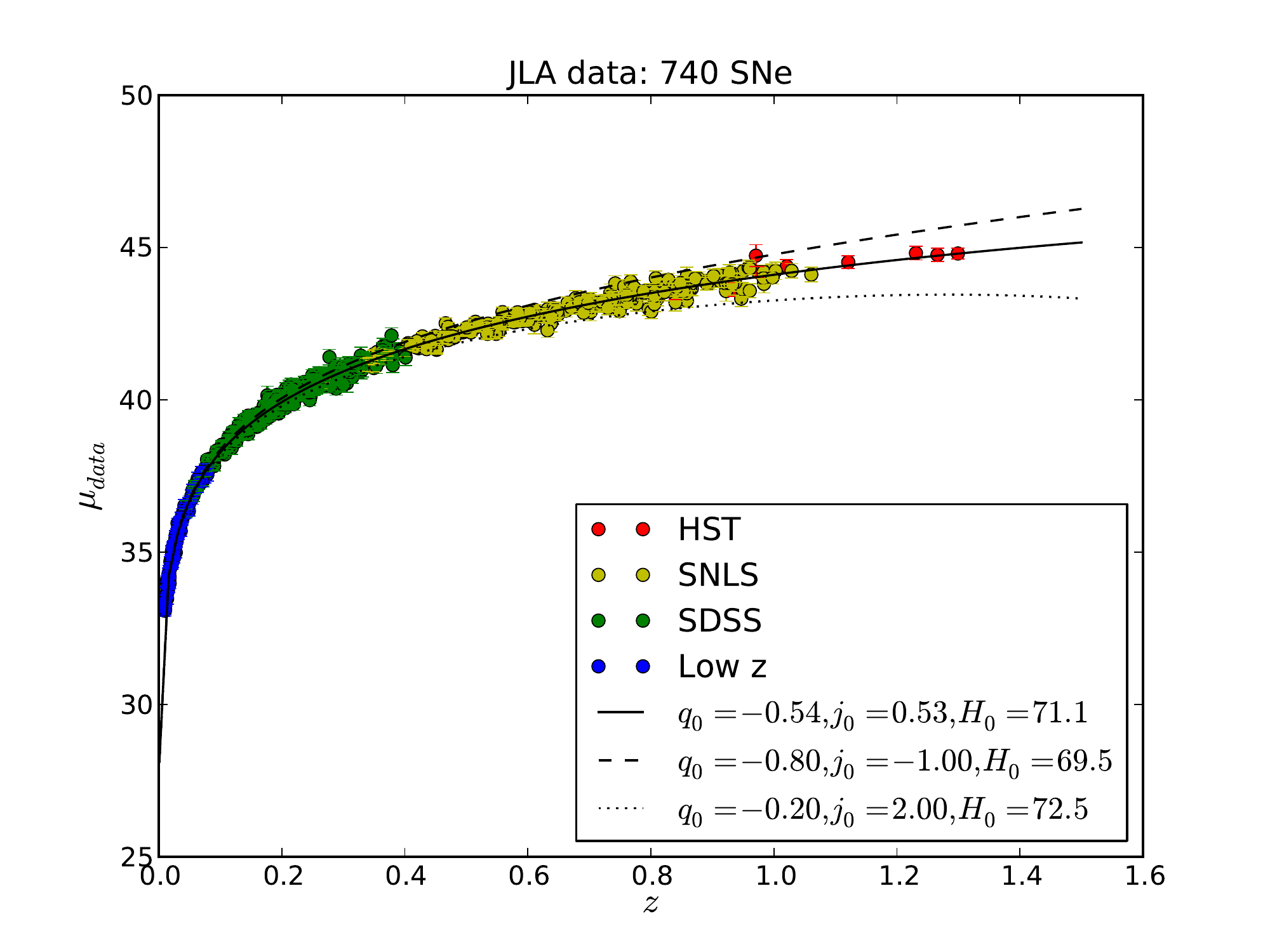}
\includegraphics[width=9cm]
%{chain_cosmic_var_fix_k_mcmc_q0j0sig0_h0sig1_ichain9_nsim10000_q0j0_h0coljet.pdf}
{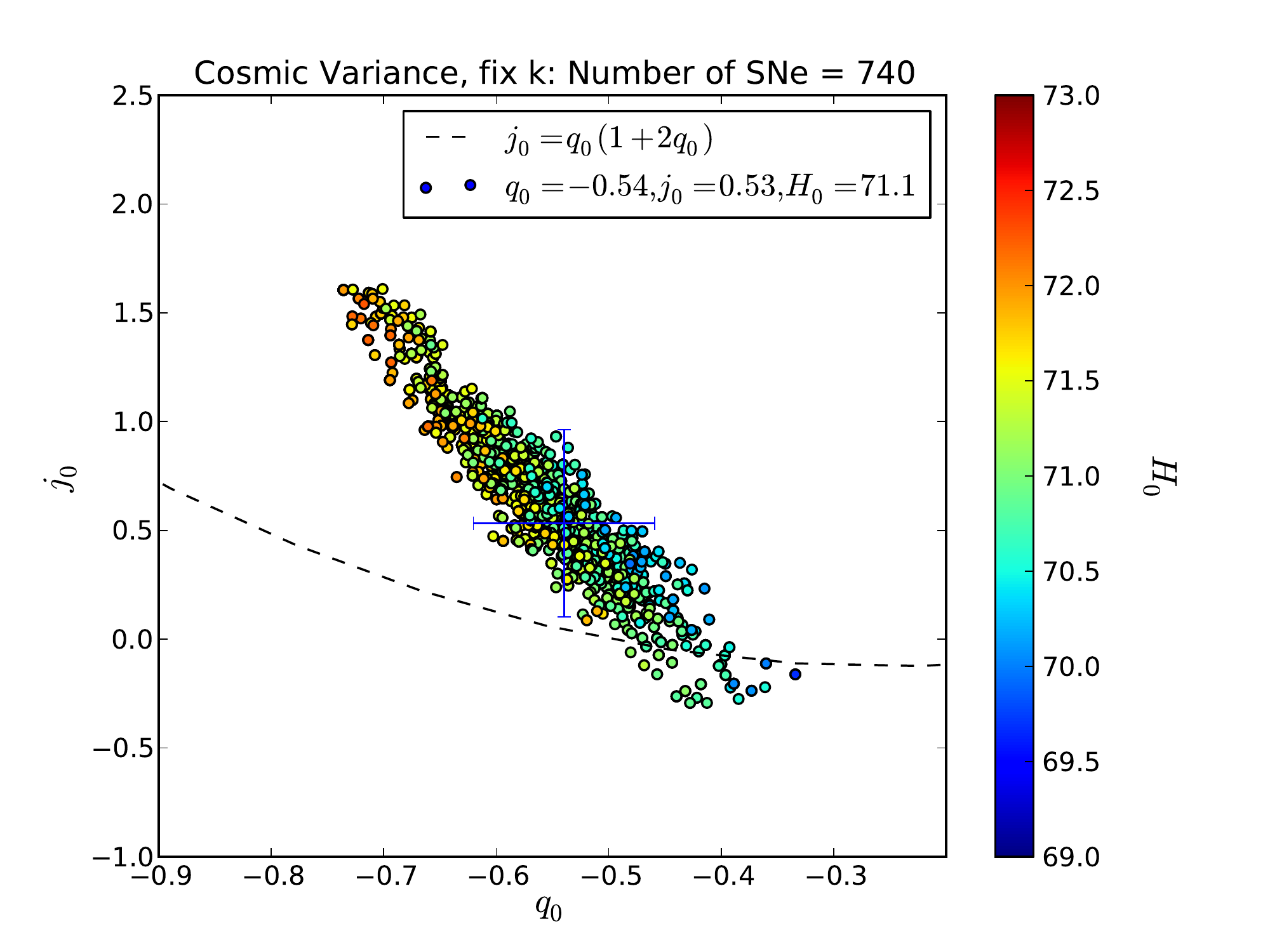}
}
\caption{%\baselineskip=0.5cm{
{\bf Left panel: distance modulus as a function of redshift.} The dots are coloured according to the original SN survey, as described in~\citet{betoule_2014}, and indicate the distance modulus $\mu_{\text{data}}$ computed from each SN in the sample. Blue: Several low--z. Green: SDSS--II. Yellow: SNLS. Red: HST. The black line indicates the distance modulus computed theoretically for the values $\{H_0,q_0,j_0\}$ estimated using the complete sample. The black dashed and black dotted lines indicate the distance modulus for two extreme cases.
{\bf Right panel: Markov chain of one realization.} The dots indicate the entries in the Markov chain. The position in the graph indicates the values of $(q_0,j_0)$ of each entry according to the axes, whereas the colour indicates the value of $H_0$ as detailed in the side colour bar. The dashed line represents $j_0=q_0(1+2q_0),$ which separates between flat and curved space--time.%}
}
\label{fig:mcmc}
%\end{figure}
\end{figure*}

For illustration, in the left panel of Fig.~\ref{fig:mcmc} we plot $\mu_{\text{theo}}$ computed for the estimated values $x_{i}^{\text{fid}}$ as the solid black line. In the right panel we plot the Markov chain of one realization. 

We recall that \citet{riess_2004} found  $q_0\in [-1.3,-0.2]$ (using the gold sample) which implies $j_0\in [-0.3,5.9],$ and $q_0\in [-1.4,-0.3]$ (using the combined gold+silver sample) which implies $j_0\in [-0.1,6.4].$ Moreover, \citet{caldwell_2004} found $q_0\in [-1.1,-0.2]$ and $j_0\in [-0.5,3.9].$ More recent results include Riess et al.'s $H_0=73.8\pm2.4$ for $q_0=-0.55$ and $j_0=1$ using the Hubble Space Telescope set \citep{riess_2011}, and Neben \& Turner's $q_0=-0.64\pm 0.14$ and $j_0=1.4\pm 0.8$ using the Constitution set \citep{neben_2013}. (See Table~\ref{table:param}.)

%\begin{figure}

\subsection{Local parameter estimation}

We then divide the SNe over a pixelated map of the sky with pixels of equal surface area according to the HEALPix pixelation \citep{gorski_2005}, and perform a parameter estimation using the SN subsample that falls into each pixel, called the local estimation. The number of pixels that guarantees non--empty SN subsamples in all pixels is ${\tt npixel}=12$ pixels. The number of SNe in each pixel is indicated in 
Fig.~3 in \citet{carvalho_2015}.
Each pixel is assumed to be described by a Friedmann--Lema\^itre--Roberston--Walker metric so that the full sky is an inhomogeneous ensemble of disjoint, locally homogeneous regions.

For the local estimation, we distinguish two cases as introduced in \citet{carvalho_2015}: the ``Cosmic variance'' estimation where in each subsample we use the original redshifts and positions, and the ``Shuffle SNe'' estimation where in each subsample we randomly shuffle the SNe in redshift while keeping the original positions in the sky. 
The ``Shuffle SNe'' estimation is hypothesised to be a measure of the noise bias due to the inhomogeneous coverage of the sky by the SN surveys, from which there results inhomogeneity in the SN subsampling. We model this noise bias by the local estimation obtained from randomizing the dependence of redshift with position, while keeping the original SN positions in the sky, and then averaging over the various randomizations.

In each pixel $k$ and for each case of local estimation, we ran 
$n_j=300$
realizations of Markov chains of length $10^{4}$ 
and repeated the procedure described above for the global estimation. From the local estimation, there result maps for each estimated parameter $x_i$ with the same pixelation as the SN subsamples, 
denoted by $\bar x_{ik}=\big<x_{ijk}\big>_j$ in the ``Cosmic variance'' estimation
and by $\bar x_{ik}^{\text{bias}}=\big<x_{ijk}^{\text{bias}}\big>_j$ in the ``Shuffle SNe'' estimation, where the brackets denote averaging over the $j$ realizations.
Subtracting the noise bias $\bar x_{ik}^{\text{bias}}$ off $\bar x_{ik},$ we obtain unbiased maps $\bar x_{ik}^{\text{unbias}}\equiv \bar x_{ik}-\bar x_{ik}^{\text{bias}}+x_i^{\text{fid}}.$ %where $\bar x_{ik}^{\text{bias}}\equiv\big< x_{ij^{\prime}k}^{\text{bias}}\big>_{j^{\prime}}$ is the average over $j^{\prime}$ randomizations in $z.$ (Here, $j^{\prime}=j.$) 
For convenience, we also compare the local estimation with the fiducial values by defining the difference maps $\Delta \bar x_{ik}=\bar x_{ik}-x_{i}^{\text{fid}},$ and the unbiased difference maps as $\Delta \bar x_{ik}^{\text{unbias}}\equiv \bar x_{ik}^{\text{unbias}}-x_{i}^{\text{fid}}=\bar x_{ik}-\bar x_{ik}^{\text{bias}}.$

%[Compare dispersion about fiducial values: error of local estimation]

\begin{figure*}
\vspace{-0.2cm}
\centerline{
\includegraphics[width=9cm]%{shift_pix_cosmic_var_fix_k_corr_sn_per_pix_ndim3_ishift0_nchain300_nsim10000_ylim2.pdf}
{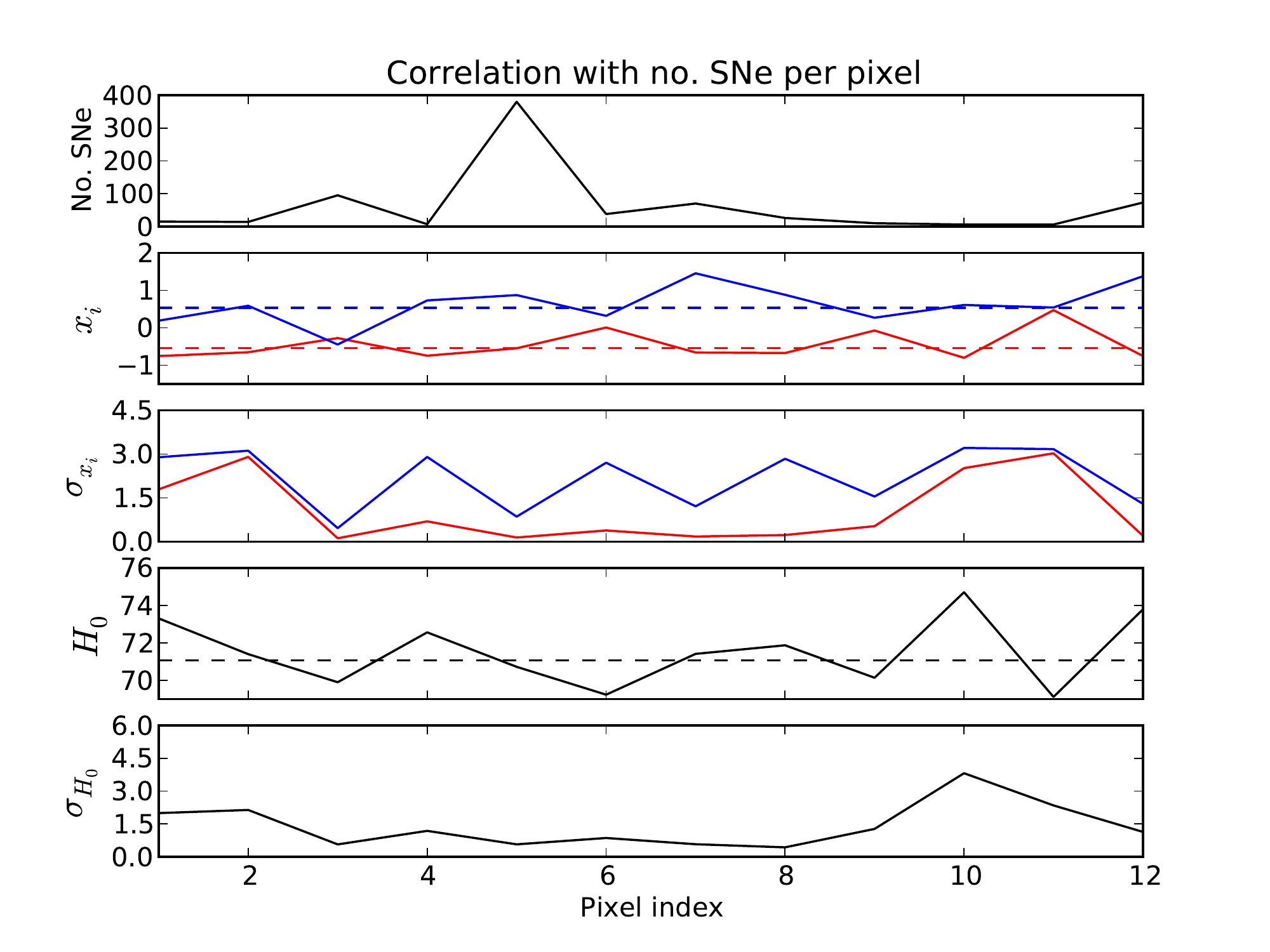}
\hspace{-1cm}
\includegraphics[width=9cm]%{shift_pix_cosmic_var_fix_k_unbias_map_corr_sn_per_pix_ndim3_ishift0_nchain300_nchainbias300_nsim10000_ylim2.pdf}
{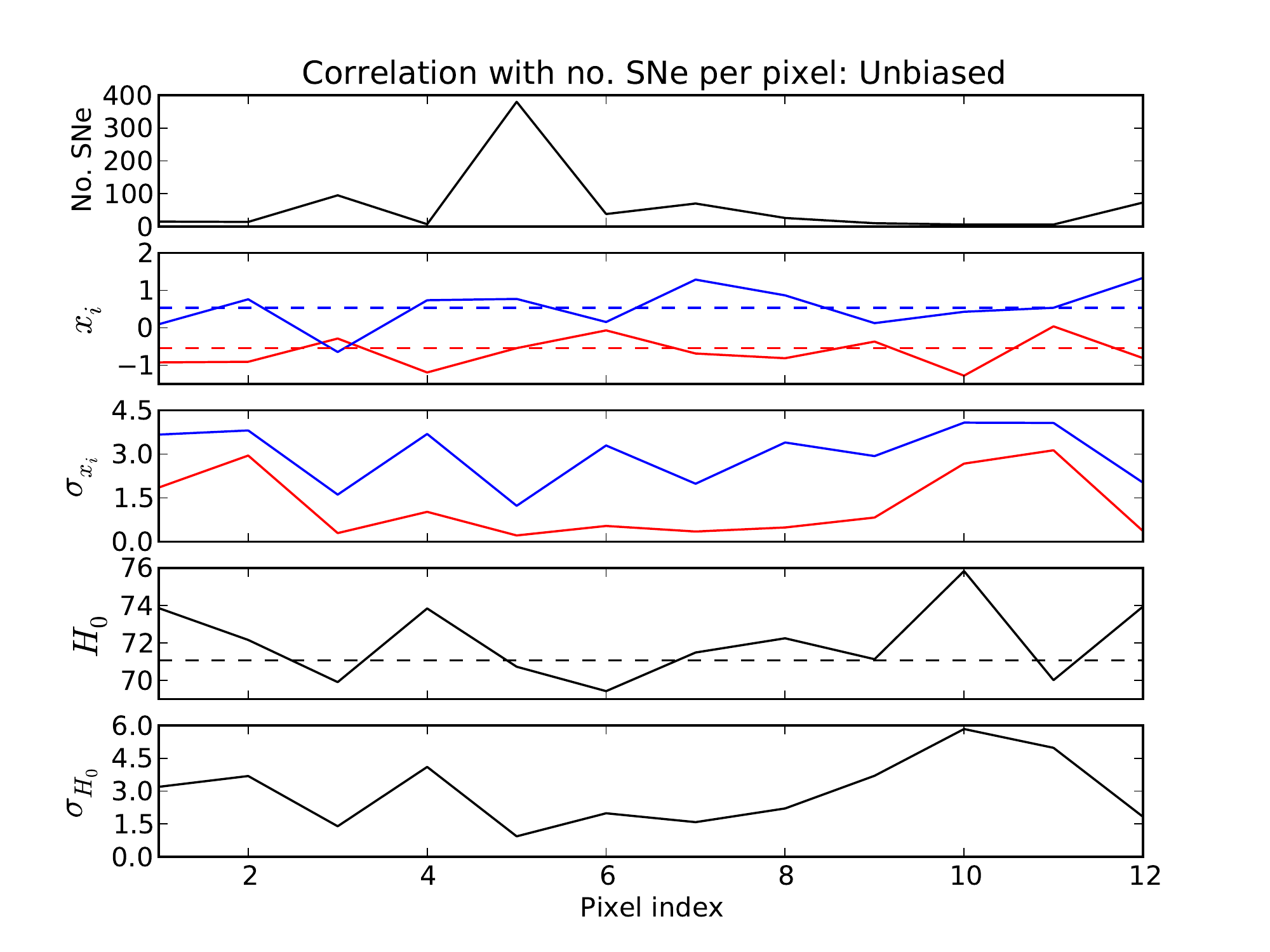}
}
\vspace{-0.2cm}
\centerline{
\includegraphics[width=9cm]%{shift_pix_cosmic_var_fix_k_corr_sn_per_pix_ndim3_nshift9_nchain300_nsim10000_ylim2.pdf}
{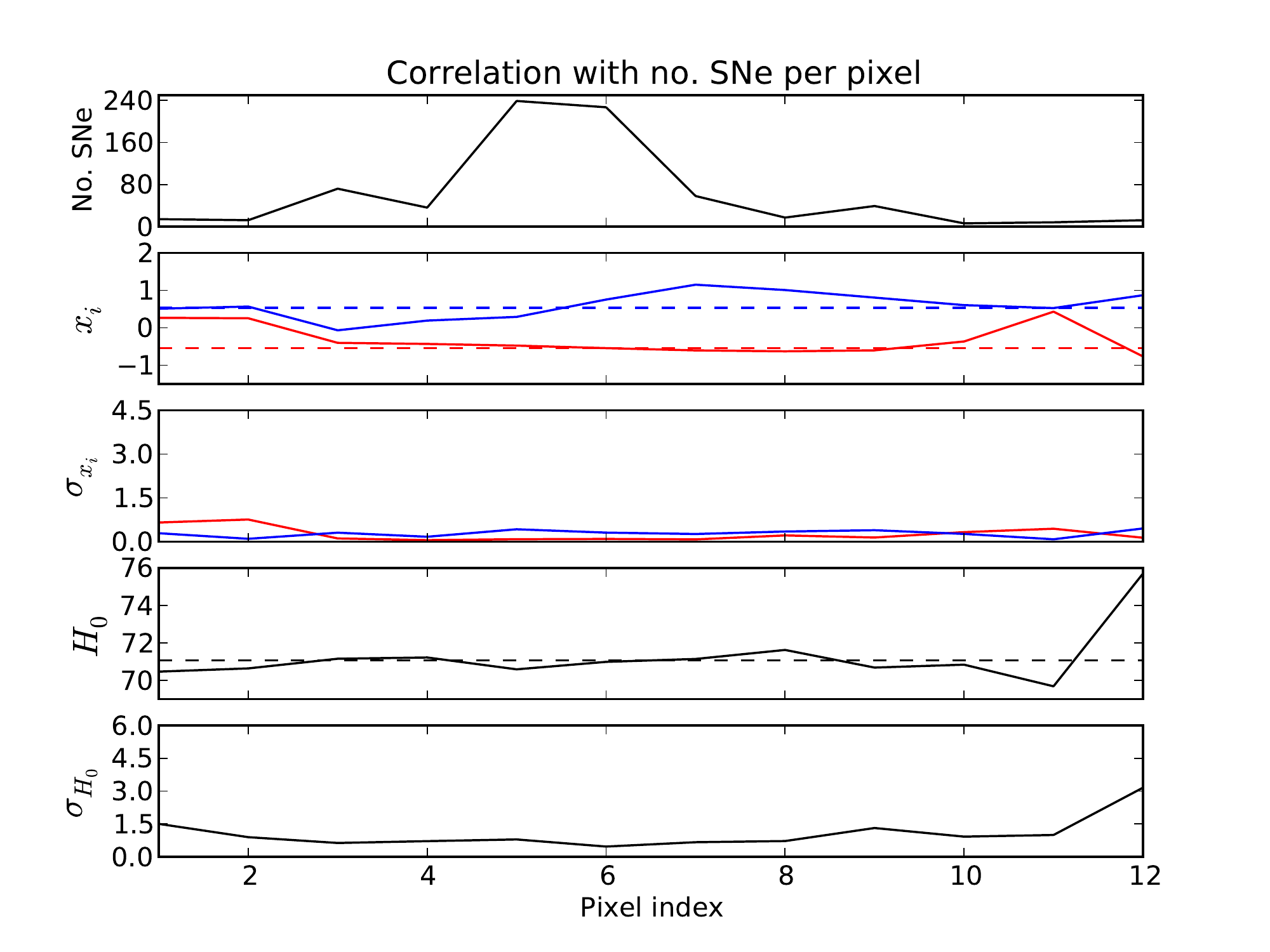}
\hspace{-1cm}
\includegraphics[width=9cm]%{shift_pix_cosmic_var_fix_k_unbias_map_corr_sn_per_pix_ndim3_nshift9_nchain300_nchainbias300_nsim10000_ylim2.pdf}
{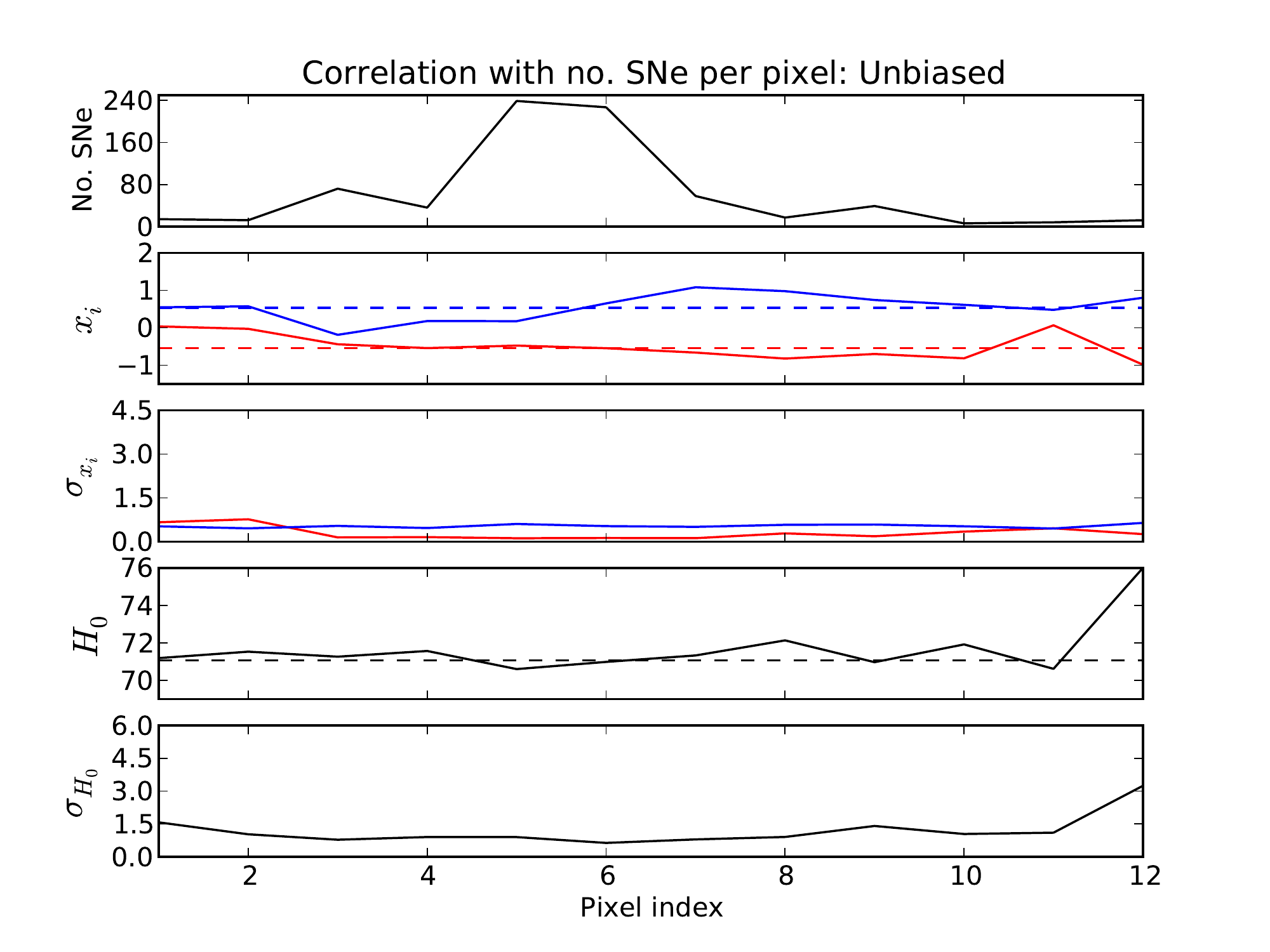}
}
\vspace{-0.2cm}
\centerline{
\includegraphics[width=9cm]
%{cosmic_var_corr_sn_per_pix_ndim3_nchain30_ylim2_2.pdf}
{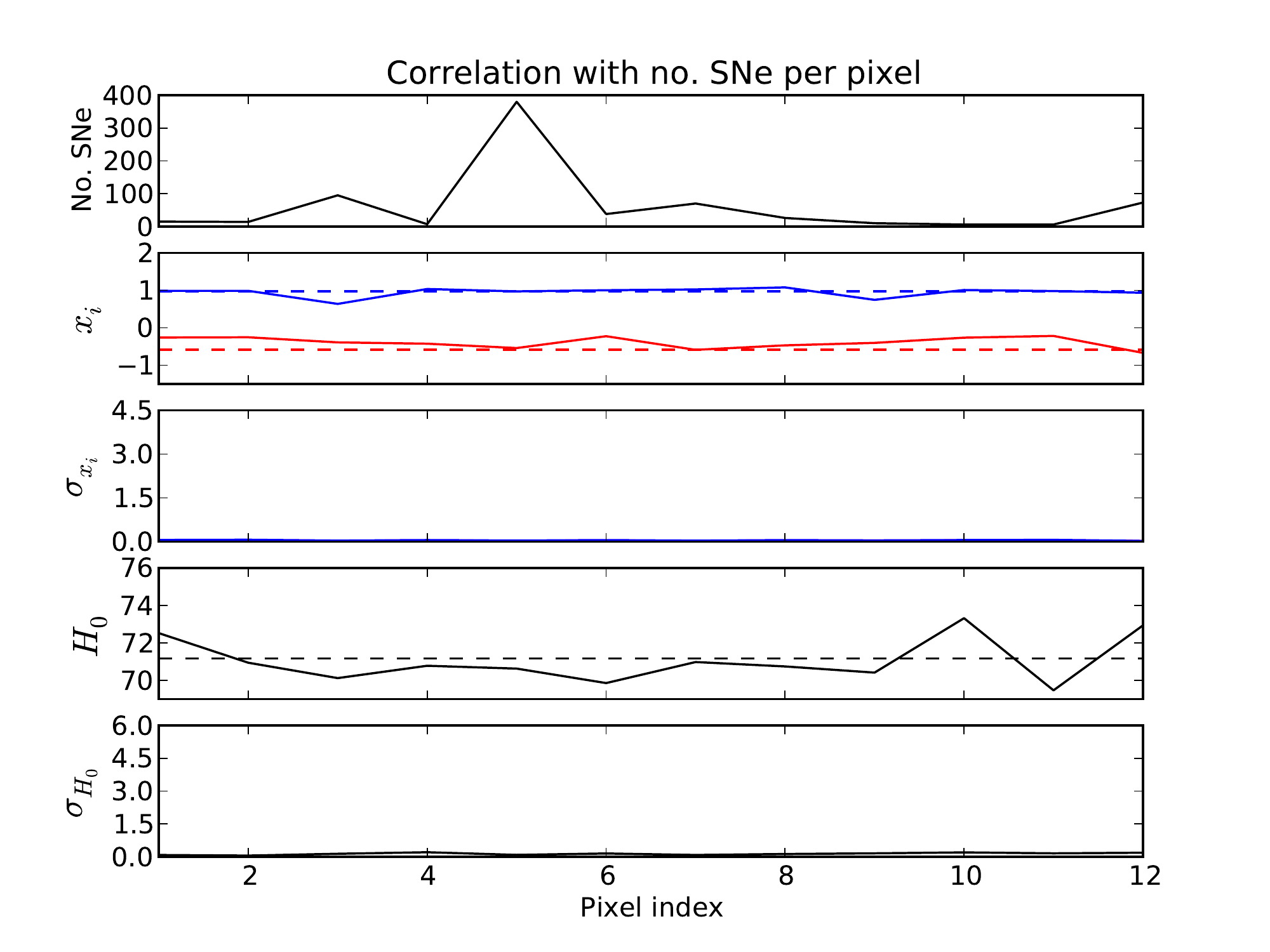}
\hspace{-1cm}
\includegraphics[width=9cm]
%{cosmic_var_unbias_map_corr_sn_per_pix_ndim3_nchain30_nchainbias100_ylim2_2.pdf}
{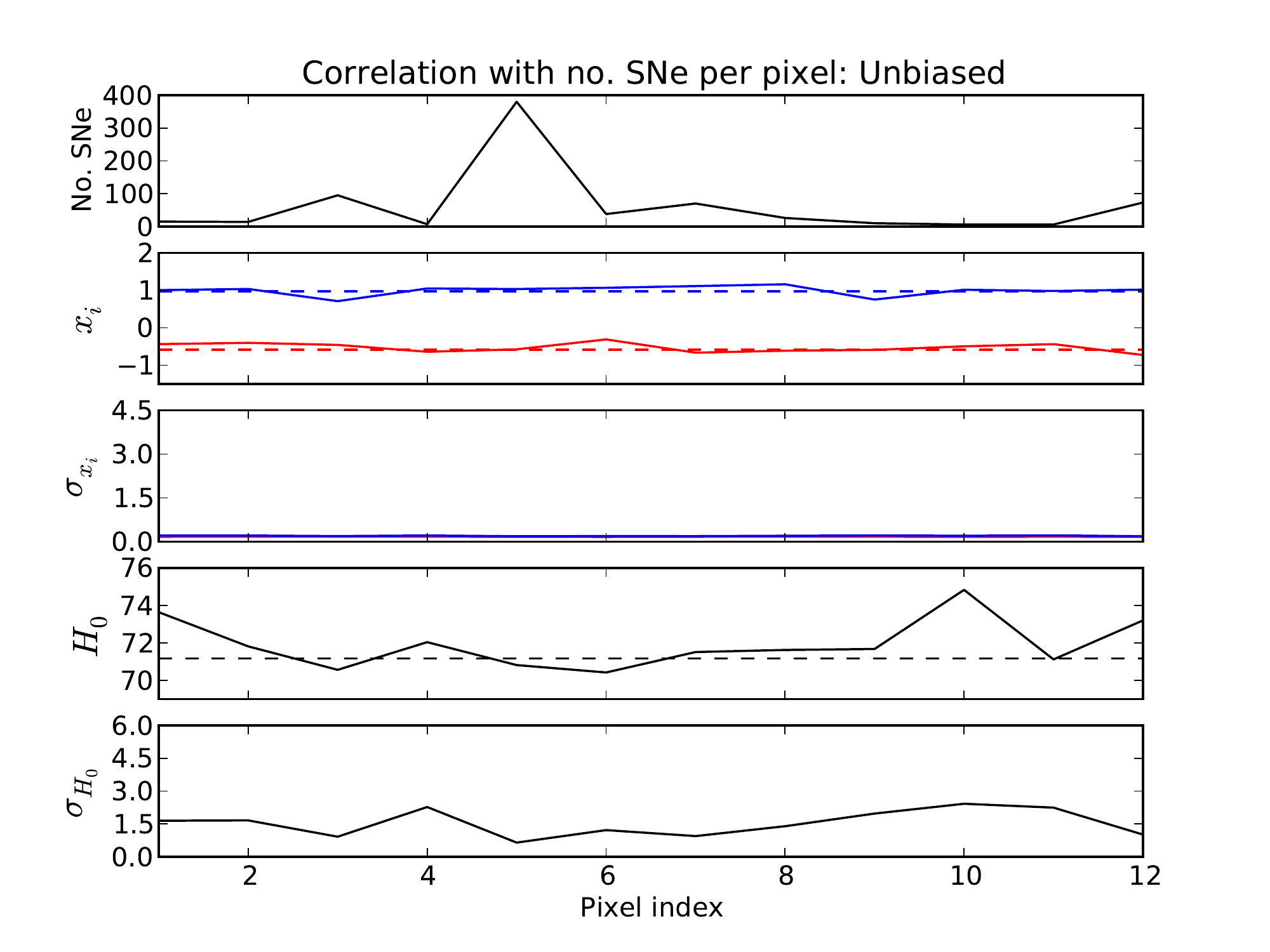}
}
\vspace{-0.2cm}
\caption{%\baselineskip=0.4cm{
{%Correlation of the number of SNe per pixel with the estimated parameters' fluctuations, 
\bf Fluctuations of the estimated parameters as a function of the number of SNe per pixel,
before (left panel sets) and after (right panel sets) noise bias removal.} Each panel set consists of five panels. At each pixel we plot: a) in the first panel, the number of SNe; b) in the second panel, the value of the parameters $\{q_0, j_0\}$ respectively as the solid red and solid blue lines, with the dashed red and dashed blue lines indicating the corresponding fiducial values; c) in the third panel, the standard deviation $\{\sigma_{q_0},\sigma_{j_0}\}$ respectively as solid red and solid blue lines; d) in the fourth plot, the value of $H_{0}$ as the solid black line, with the dashed black line marking the fiducial value; e) in the fifth panel, the standard deviation $\sigma_{H_{0}}$ as the solid black line. Top panel sets: results from the $\{H_{0},q_0,j_{0}\}$ estimation for one pixelation. Centre panel sets: results from the $\{H_{0},q_0,j_{0}\}$ estimation for the average over the different pixelations. Bottom panel sets: results from the $\{H_{0},\Omega_{M},\Omega_{\Lambda}\}$ estimation for $\Omega_{\kappa}=1-\Omega_{M}-\Omega_{\Lambda}.$ %free.%}
}
\label{fig:corr_per_pix_shift}
%\end{figure}
\end{figure*}

The difference maps 
are shown in Fig.~\ref{fig:corr_per_pix_shift} (top panel sets), before (left panel set) and after (right panel set) the noise bias subtraction.
Comparing the difference maps with the fiducial values, we measure fluctuations of order 
0.1--5\% 
for $H_{0}$, 
1--187\% 
for $q_{0}$ and 
1--184\% 
for $j_{0}$ before the noise bias subtraction; after the noise bias subtraction, we measure fluctuations of order 
0.1--7\% 
for $H_{0}$, 
1--136\% 
for $q_{0}$, and 
1--221\% 
for $j_{0}.$ 
%{\color{red}The noise bias removal brings the pixel values closer to the corresponding values estimated from the complete sample, albeit with a boost in the amplitude of the fluctuations in the case of $j_{0}.$ Simultaneously, it increases the error by 
%a factor of up to 
%$\{5.,2.5,3.5\}$ respectively.}
%(or up to 
%$\{410,153,248\}\%$ %[CHECK] %ishift=0,nchain=300,nchain_bias=300
%respectively).

%[Comparative analysis]
We recall that \citet{wiegand_2012} found $\delta H_0/H_0\ltorder 5\%$ using galaxy surveys, which is consistent with our results.

For comparison, in the bottom panels of Fig.~\ref{fig:corr_per_pix_shift}, we reproduce the difference maps of $\{H_{0},q_0=\Omega_{M}/2-\Omega_{\Lambda},j_{0}=\Omega_{M}+\Omega_{\Lambda}\}$ from the $\{H_0, \Omega_{M},\Omega_{\Lambda}\}$ estimation in \citet{carvalho_2015}.
We recall that the $\{H_{0},\Omega_{M},\Omega_{\Lambda}\}$ estimation yielded fluctuations about the fiducial values of order 
0.1--3\% 
for $H_{0}$, 
0.1--63\% 
for $q_{0}$, and 
0.001--34\% 
for $j_{0}$ before the noise bias subtraction, and fluctuations about the fiducial values of order 
0.1--5\% 
for $H_{0}$, 0.1--32\% 
for $q_{0}$, and 
1--27\% 
for $j_{0}$ after the noise bias subtraction. This amounts to an increase in the fluctuations by a factor of 
four and two orders of magnitude 
respectively before and after the noise bias subtraction, between the $\{H_0, \Omega_{M},\Omega_{\Lambda}\}$ and the $\{H_{0},q_0,j_{0}\}$ estimations. 
The increase in the fluctuations supports the reasoning 
that accuracy in the estimation of $\{H_0,q_0\}$ is compromised by the inclusion of $j_0$ as an estimated parameter. However, this is also a consequence of using a parametrisation that is independent of the cosmological equation of state, hence of assuming less about the cause of acceleration. 

For further comparison, we compute the largest fluctuation in the maps, defined as $\text{max}(\Delta \bar x_{ik})\equiv \text{max}(\bar x_{ik})-\text{min}(\bar x_{ik}),$ which yields 
$\text{max}(\{\Delta H_{0},\Delta q_{0},\Delta j_{0}\})=\{5.58,1.27,1.90\}$
and $\text{max}(\{\Delta H_{0},\Delta q_{0},\Delta j_{0}\})=\{6.40,1.31,1.97\},$ before and after the noise bias subtraction respectively.
We recall that the $\{H_{0},\Omega_{M},\Omega_{\Lambda}\}$ estimation yielded 
$\text{max}(\{\Delta H_{0},\Delta q_{0},\Delta j_{0}\})=\{3.84,0.452,0.440\}$ 
and  $\text{max}(\{\Delta H_{0},\Delta q_{0},\Delta j_{0}\})=\{4.40,0.416,0.450\},$ before and after the noise bias subtraction respectively. 
In comparison with the results from other SN studies, namely  
Kalus et al.'s $2.0<\text{max}(\Delta H_{0})<3.4~\text{km~s}^{-1}\text{Mpc}^{-1}$ from the Union 2 data for fixed $q_{0}=-0.601$ \citep{kalus_2013}, and 
Bengaly et al.'s $\text{max}(\Delta H_{0})=\{4.6, 5.6\}~\text{km~s}^{-1}\text{Mpc}^{-1}$ and $\text{max}(\Delta q_{0})=\{2.56, 1.62\}$ respectively from the Union 2.1 data and the JLA data \citep{bengaly_2015}, we observe that: a) our results for $H_{0}$ remain intermediate between the previous results, b) our results for $q_{0}$ are intermediate between those of \cite{bengaly_2015} for the two data sets, and c) our results for $j_{0}$ are to the best of our knowledge the first measurements of the kind.

%[TO DO: COMMENT ON THE ERRORS DUE TO INHOMOGENEOUS COVERAGE BEING LARGER THAN PARAMETER UNCERTAINTIES, HENCE THAN CORRELATED ERRORS] \citep{neben_2013}

%[Correlations with galactic plane]

%Although the Galactic plane could contribute a spurious inhomogeneity in the local estimation, 
A visual inspection of the pixels through which the Galactic plane crosses does not reveal consistently larger/smaller values of the fluctuations, as also noted in \citet{carvalho_2015}, thus suggesting negligible correlation with the Galactic plane in comparison with the measurement errors \citep{neben_2013}.

%\begin{figure}

%\subsection{Local parameter estimation upon rotating the SN subsampling}\subsection{Inhomogeneity test: Rotate the SN subsampling per pixel}

In order to verify the dependence of the results on the subsampling of the SN surveys per pixel, we perform a test of the effect of rotating the HEALPix pixelation. In particular, starting from the current pixelation and keeping the pixel size constant (i.e. keeping the same number of pixels on the sky), we consider $n$ divisions of each pixel and rotate the pixelation $(n_p-1)$ times in the same direction, the $n_p$th rotation corresponding to the initial pixelation. At each such rotation by a fraction of $1/n_p$ of the pixel, we obtain 
a different pixelation of the sky and hence a different subsampling of the SN surveys per pixel, totalling $n_p$ different pixelations. We choose $n_p=9$ to guarantee a sufficient number of rotations and simultaneously little degeneracy among the rotations.

For each such pixelation $p,$ we perform a local parameter estimation as described above, consisting of a ``Cosmic covariance'' estimation and a ``Shuffle SNe'' estimation, and hence obtaining an unbiased parameter estimation in each pixel $k,$ $\bar x^{\text{unbias}}_{ikp}.$ 
We then compute the mean unbiased maps averaged over the different pixelations as $\bar x_{ik}^{\text{unbias}}=\big<\bar x_{ikp}^{\text{unbias}}\big>,$ weighted by the 
%number of SNe in each pixel (the mean pixel derived from the rotation of the same pixel in the initial pixelation) 
inverse of the variance in each pixel, 
over the different pixelations. 
From the average over the different pixelations there results a mean pixel derived by averaging the $(n-1)$ rotations of the same pixel in the initial pixelation.
For convenience, we define the unbiased difference maps as $\Delta \bar x_{ik}^{\text{unbias}}\equiv \bar x_{ik}^{\text{unbias}}-x_{i}^{\text{fid}}.$

The resulting difference maps are shown in Fig.~\ref{fig:corr_per_pix_shift} centre panel sets, before (left panel set) and after (right panel set) the noise bias subtraction.
Comparing the difference maps with the fiducial values, we measure fluctuations of order 
0.1--7\% 
for $H_0,$ 
0.1--180\% 
for $q_0,$ and 1--116\% 
for $j_0$ before the noise bias subtraction; after the noise bias subtraction we measure fluctuations of order 
0.1--7\% 
for $H_0,$
0.1--112\% 
for $q_0,$ and 
1--135\% 
for $j_0.$
The noise bias removal brings the pixel values closer to the corresponding values estimated from the complete sample, hence decreasing the fluctuations across the sky, albeit increasing slightly the fluctuations in $j_0.$ Simultaneously, it increases the error by 
up to 
$\{10,65,5\}\%$ 
respectively. 
These results also seem to indicate the validity of $\bar x_{ik}^{\text{bias}}$ as a measure of the noise bias due to the inhomogeneous SN sampling.

Similarly, the largest fluctuation in the maps yields 
$\text{max}(\{\Delta H_{0},\Delta q_{0},\Delta j_{0}\})=\{6.01,1.19,1.22\}$
and $\text{max}(\{\Delta H_{0},\Delta q_{0},\Delta j_{0}\})=\{5.40,1.05,1.27\},$ before and after the noise bias subtraction.

Hence by averaging over different pixelations of the sample, we obtain a decrease in the fluctuations of the parameters across the sky.
However, the mean fluctuations are still larger (by a factor of four and two orders of magnitude 
respectively before and after the noise bias subtraction) than those obtained in the $\{H_0, \Omega_{M},\Omega_{\Lambda}\}$ estimation. 
%{\color{red}Given the larger angular fluctuations about the fiducial values, we expect larger angular correlations and larger backreaction effects.}
In the subsequent calculations, we will use the parameters' maps obtained from all $n_p$ pixelations.

\section{Power spectra}
\label{sec:power}

In order to probe the distribution of the fluctuations by scale, we compute the angular power spectrum of the difference maps normalised to the corresponding fiducial parameter value. The equations below follow closely those in \citet{carvalho_2015} except for the extra degree of complexity introduced by the different pixelations. In particular, for each parameter $x_i$ and for each realization $j$ of the ``Cosmic variance'' local estimation, we define the map with value $\delta x_{ijkp}\equiv (x_{ijkp}-x^{\text{fid}}_{i})/x^{\text{fid}}_{i}$ at each pixel $k,$ for each pixelation $p.$ An estimator of the power spectrum is
\ba
\hat C_{\ell,ijp}={1\over{(2\ell+1)}}\sum_{m=-\ell}^{\ell}\vert \hat a_{\ell m,ijp}\vert^2,
\ea
where the harmonic coefficients $\hat a_{\ell m,ijp}$ for a pixelated map of $N_{\text{pix}}$ pixels are given by
\ba
\hat a_{\ell m,ijp}={4\pi\over N_{\text{pix}}}\sum_{k=1}^{N_{\text{pix}}}\delta x_{ijkp}Y^{\ast}_{\ell m}(\boldsymbol{n}_k).
\ea
We compute the mean power spectrum $\hat C_{\ell,ip}$ for the parameters $\{H_0,q_0,j_0\}$ by averaging the power spectra $\hat C_{\ell,ijp}$ of the maps $\delta x_{ijkp}=\{\delta H_0,\delta q_0, \delta j_0\}_{jkp,\text{cosmic\_var}}$ over the $j$ realizations of the ``Cosmic variance'' local estimation
\ba
\hat C_{\ell,ip}=\big<\hat C_{\ell,ijp}\big>_j,
\ea
with variance
\ba
\text{Var}[\hat C_{\ell,ip}]=\text{Var}[\hat C_{\ell,ip}]^{\text{sample}}+\text{Var}[\hat C_{\ell,ip}]^{\text{estimator}},
\ea
where
\ba
\text{Var}[\hat C_{\ell,ip}]^{\text{estimator}}={2\over {2\ell+1}}\hat C_{\ell,ip}^2.
\ea
We then compute the mean power spectrum $\hat C_{\ell,i}$ over the $p$ pixelations 
\ba
\hat C_{\ell,i}=\big<\hat C_{\ell,ip}\big>_p,
\ea
with variance given by 
\ba
\text{Var}[\hat C_{\ell,i}]=\left[\sum_{p=1}^n {1\over \text{Var}[\hat C_{\ell,ip}]}\right]^{-1}.
%\text{Var}[\hat C_{\ell,i}]={1\over \sum_{p=1}^n {1\over \text{Var}[\hat C_{\ell,ip}]}}.
\ea

In order to compute the unbiased power spectrum we devise two methods, as detailed below.

\subsection{Difference of power spectra}

\begin{figure*}
\centerline{
\includegraphics[width=9cm]%{cosmic_var_unbias_mean_cl_ldelta_mean_q0j0_omegasig0_h0sig1_nchain30_nchainbias100_nsim10000_npix12_var_total_2.pdf}
{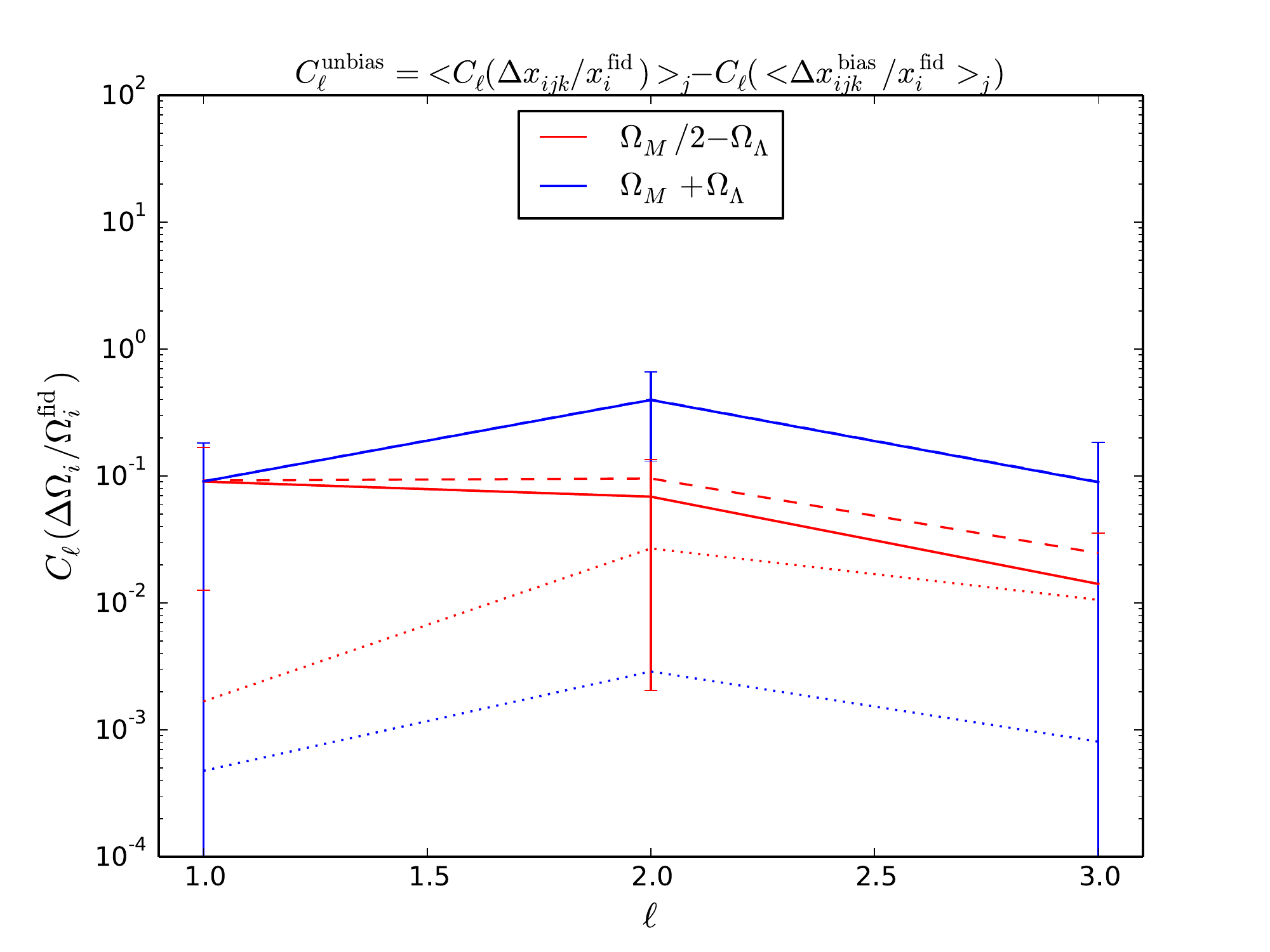}
\includegraphics[width=9cm]%{shift_pix_cosmic_var_fix_k_unbias_mean_cl_ldelta_mean_q0_q0j0sig0_h0sig1_nshift9_nchain300_nchainbias300_nsim10000_npix12_var_total.pdf}
{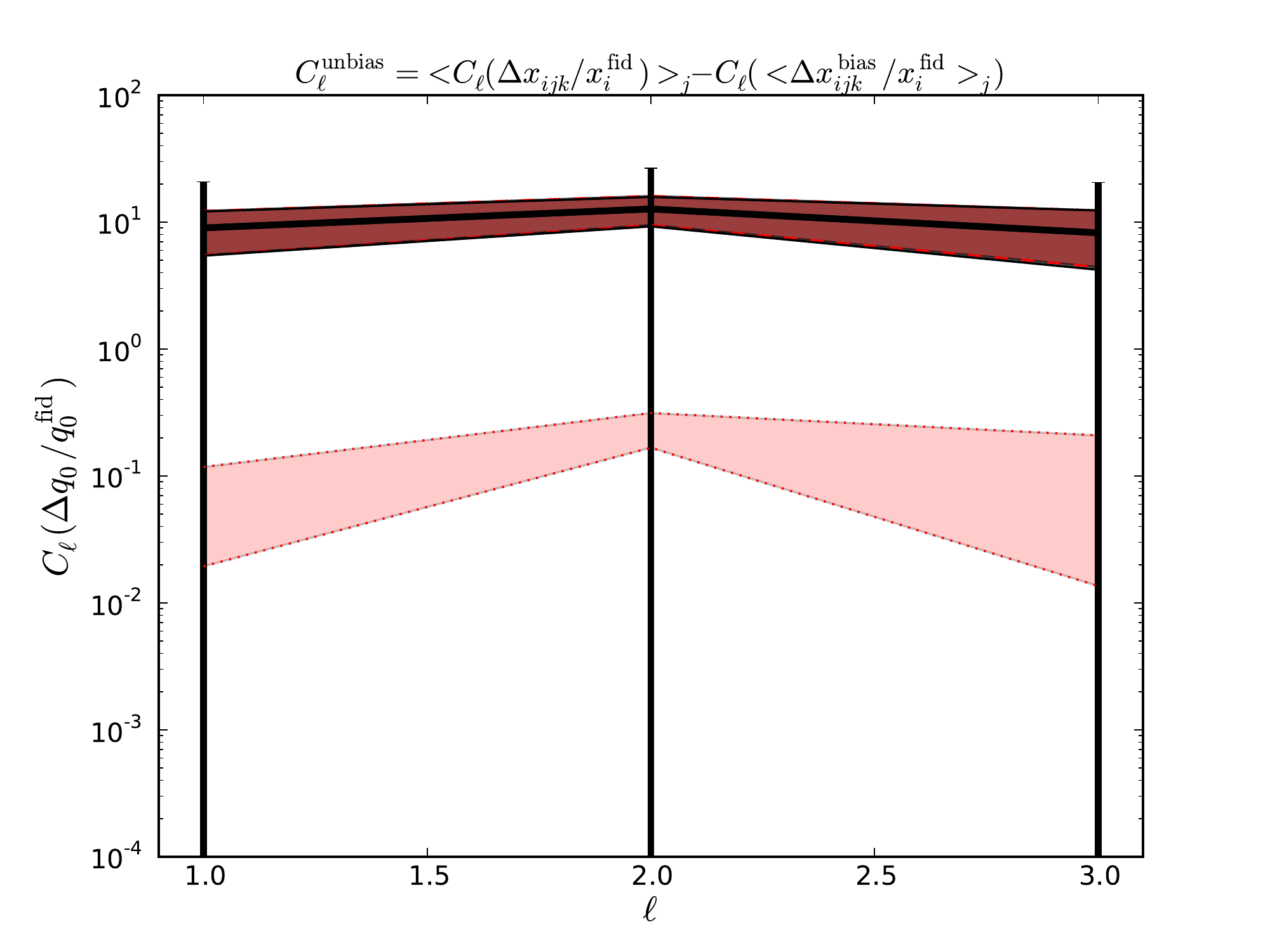}
}
\centerline{
\hskip9.1cm
\includegraphics[width=9cm]%{shift_pix_cosmic_var_fix_k_unbias_mean_cl_ldelta_mean_j0_q0j0sig0_h0sig1_nshift9_nchain300_nchainbias300_nsim10000_npix12_var_total.pdf}
{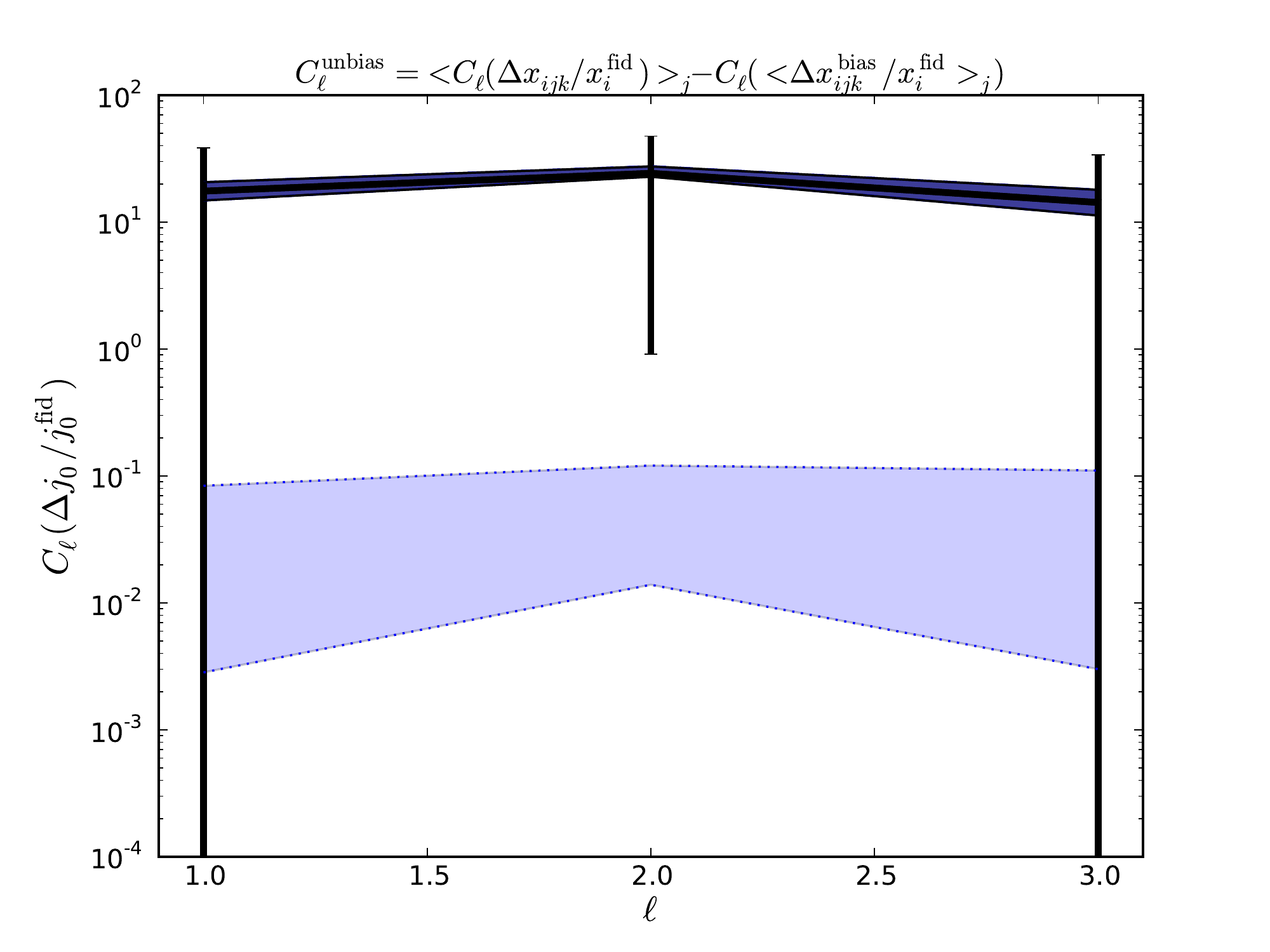}
}
\centerline{
\includegraphics[width=9cm]%{cosmic_var_unbias_mean_cl_ldelta_mean_h0_omegasig0_h0sig1_nchain30_nchainbias100_nsim10000_npix12_var_total_2.pdf}
{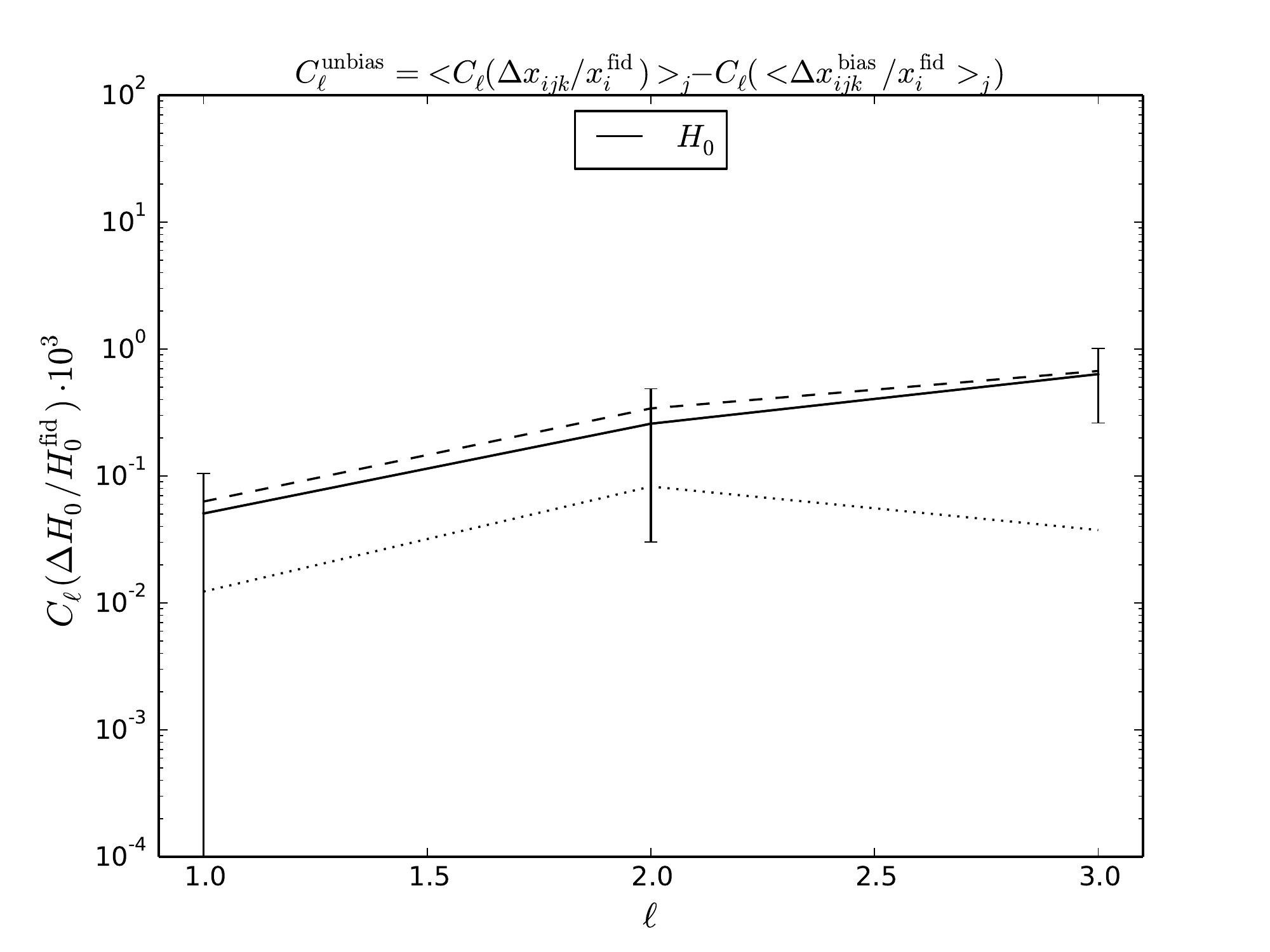}
\includegraphics[width=9cm]%{shift_pix_cosmic_var_fix_k_unbias_mean_cl_ldelta_mean_h0_q0j0sig0_h0sig1_nshift9_nchain300_nchainbias300_nsim10000_npix12_var_total.pdf}
{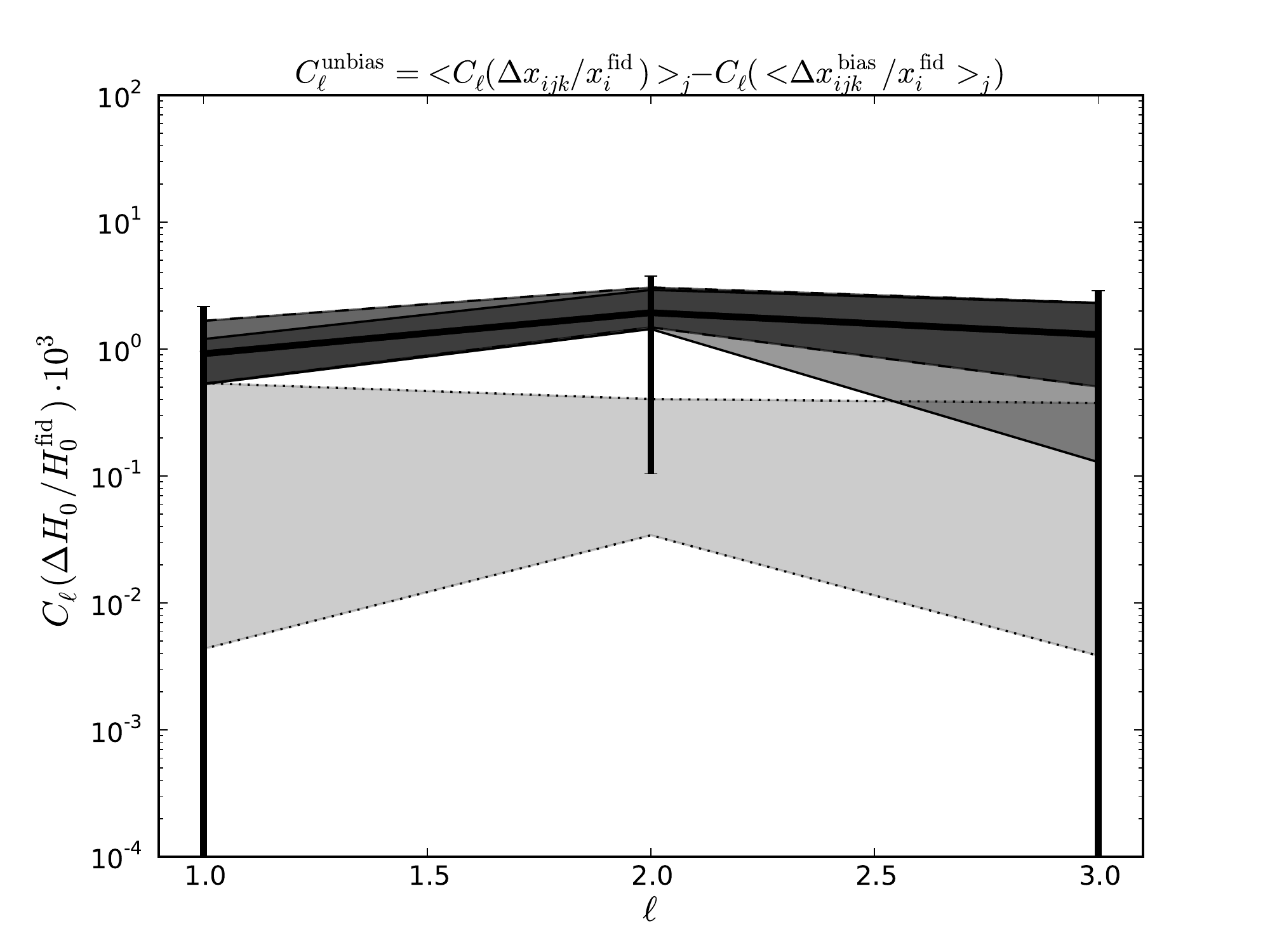}
}
\caption{%\baselineskip=0.4cm{
{\bf Power spectrum of the parameters estimated from the JLA type Ia SN sample (method A).} The dashed lines are the mean power spectra from the ``Cosmic variance'' estimation, the dotted lines are the power spectra of the average ``Shuffle SNe'' estimation and the solid lines are the unbiased power spectra defined as the difference between the former two. Left panels: Power spectra from the $\{H_0,\Omega_{M},\Omega_{\Lambda}\}$ estimation. Right panels: Power spectra from the $\{H_0,q_0,j_0\}$ estimation.%}
}
\label{fig:cl_a_shift}
%\end{figure}
\end{figure*}

In the first method, we define the unbiased power spectrum as the unbiased mean power spectrum.
We compute the unbiased mean power spectrum by computing the power spectrum $C_{\ell,ip}^{\text{bias}}$ of the mean map $\delta \bar x_{ikp}^{\text{bias}}=\big<\delta x_{ijkp}^{\text{bias}}\big>_j=\{\delta \bar H_0,\delta \bar q_0, \delta \bar j_0\}_{kp,\text{shuffle\_sn}}$ averaged over the $j$ realizations of the ``Shuffle SNe'' local estimation and subtracting it from the mean power spectrum $\hat C_{\ell,ip}$
\ba
\hat C_{\ell,ip}^{\text{unbias}}=\hat C_{\ell,ip}-C_{\ell,ip}^{\text{bias}},
\ea
with total variance
\ba
\text{Var}[\hat C_{\ell,ip}^{\text{unbias}}]
=\text{Var}[\hat C_{\ell,ip}]+\text{Var}[C_{\ell,ip}^{\text{bias}}]^{\text{estimator}}.
\ea
This was the method suggested in \citet{carvalho_2015}. We then compute the mean unbiased power spectrum $\hat C_{\ell,i}^{\text{unbias}}$ by averaging $\hat C_{\ell,ip}$ over the different $p$ pixelations
\ba
\hat C_{\ell,i}^{\text{unbias}}=\big<\hat C_{\ell,ip}^{\text{unbias}}\big>_p,
\ea
with total variance
\ba
\text{Var}[\hat C_{\ell,i}^{\text{unbias}}]=\left[\sum_{p=1}^n {1\over \text{Var}[\hat C_{\ell,ip}^{\text{unbias}}]}\right]^{-1}.
\ea
In the right panels of Fig.~\ref{fig:cl_a_shift}, we plot the resulting unbiased power spectra, for the different pixelations %(grey shaded region) 
and for the average over the $p$ pixelations. %(solid line)
The variability in $\hat C_{\ell,ip}$ and $C_{\ell,ip}^{\text{bias}}$ from the different pixelations is indicated as colour-shaded regions bordered by dashed and dotted lines respectively.
For all parameters, the power spectrum has a maximum at the quadrupole ($\ell=2$). 
However, given the size of the error, the results are also compatible with a flat spectrum.

%[Comparative analysis]

For comparison, in the left panels of Fig.~\ref{fig:cl_a_shift}, using the same line types, we plot the power spectra for the parameters $\{H_0,q_0=\Omega_{M}/2-\Omega_{\Lambda},j_0=\Omega_{M}+\Omega_{\Lambda}\}$ from the $\{H_0,\Omega_{M},\Omega_{\Lambda}\}$ estimation in \citet{carvalho_2015}. In this estimation, the unbiased power spectra also follow the same behaviour as the power spectra before the noise bias removal, with the exception of $q_0$ whose subtle maximum at $\ell=2$ is erased with the noise bias removal and the power spectrum decreases always with the multipole. Conversely, for $H_0$ the power spectrum increases always with the multipole, and for $j_0$ the power spectrum has a maximum at $\ell=2.$ 
Hence, between the $\{H_0,\Omega_{M},\Omega_{\Lambda}\}$ and the $\{H_0,q_0,j_0\}$ estimation, we observe the creation of a peak at $\ell=2$ for $H_0$ and $q_0,$ and the smoothing of the peak for $j_0.$
We also observe that the power spectra in the $\{H_0,q_0,j_0\}$ estimation are up to 
$5\times10^2$ 
times the power spectra in the $\{H_0,\Omega_{M},\Omega_{\Lambda}\}$ estimation, hence yielding an increase of the amplitude. %angular correlation.

The power spectrum that accounts for the noise bias $C_{\ell,i}^{\text{bias}}$ is up to two orders of magnitude smaller that the mean power spectra $\hat C_{\ell,i},$ the resulting unbiased mean power spectra $\hat C_{\ell,i}^{\text{unbias}}$ following the same behaviour as  
$\hat C_{\ell,i}.$ 
Since $C_{\ell,i}^{\text{bias}}\ll \hat C_{\ell,i}$ for the $\{H_0,q_0,j_0\}$ estimation, this method might not remove entirely the noise bias contribution to the power spectrum. For this reason, we conceived another method.  

%\begin{figure}

%\begin{figure}
\begin{figure*}
\centerline{
\includegraphics[width=9cm]%{cosmic_var_cl_ldelta_mean_unbias_q0j0_omegasig0_h0sig1_nchain30_nchainbias100_nsim10000_npix12_var_measure_2.pdf}
{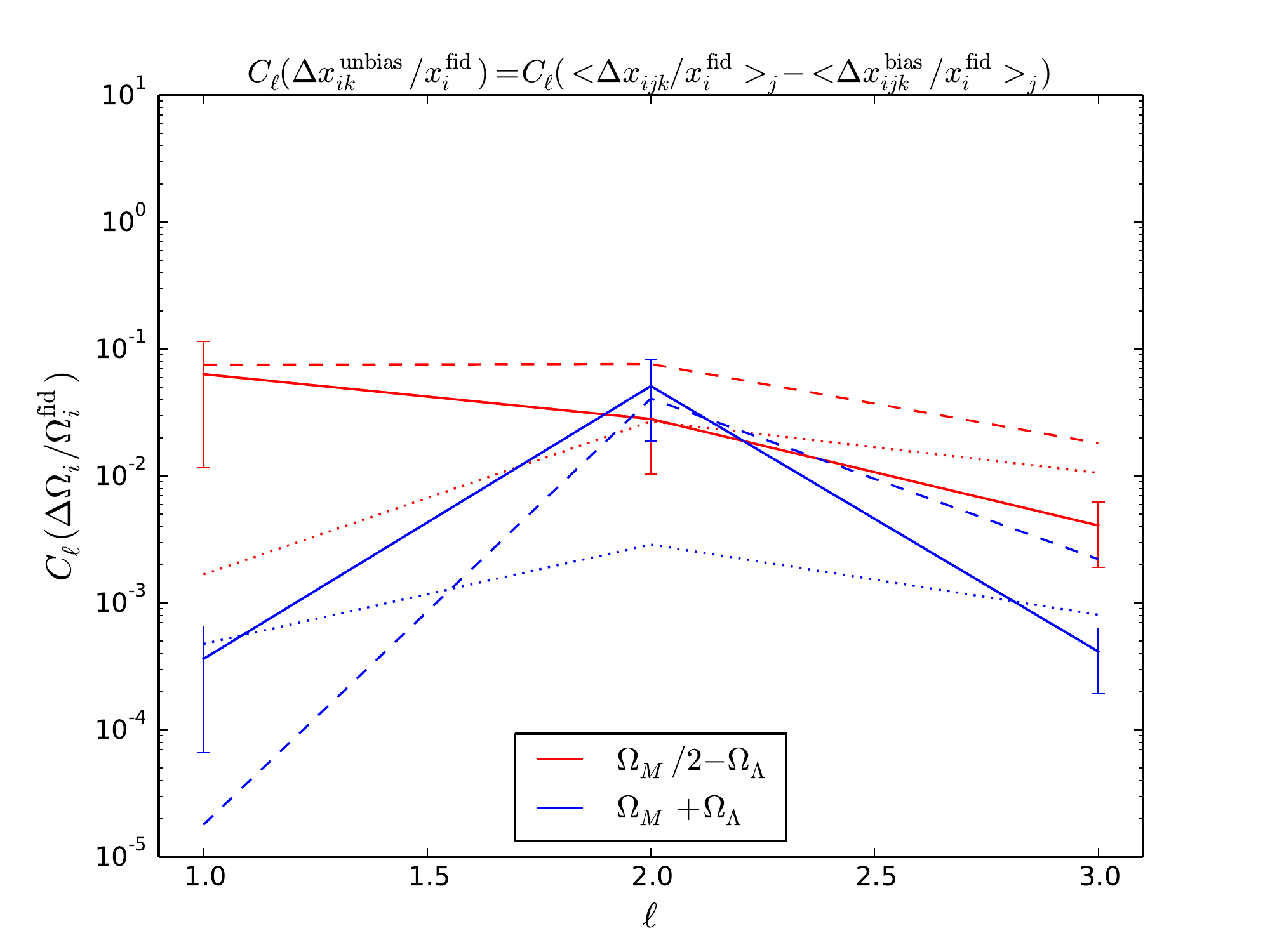}
\includegraphics[width=9cm]%{shift_pix_cosmic_var_fix_k_cl_ldelta_mean_unbias_q0_q0j0sig0_h0sig1_nshift9_nchain300_nchainbias300_nsim10000_npix12_var_measure.pdf}
{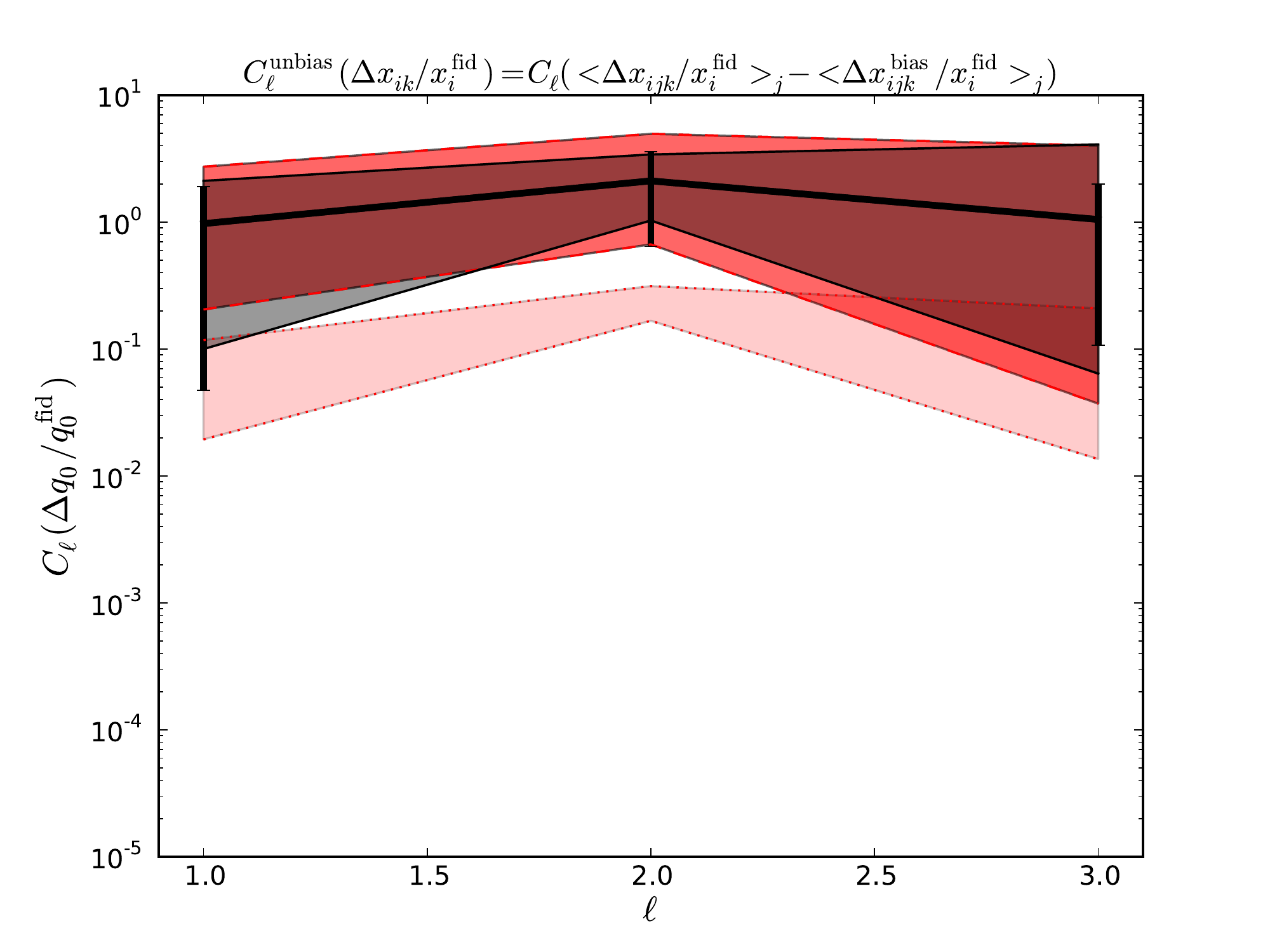}
}
\centerline{
\hskip9.1cm
\includegraphics[width=9cm]%{shift_pix_cosmic_var_fix_k_cl_ldelta_mean_unbias_j0_q0j0sig0_h0sig1_nshift9_nchain300_nchainbias300_nsim10000_npix12_var_measure.pdf}
{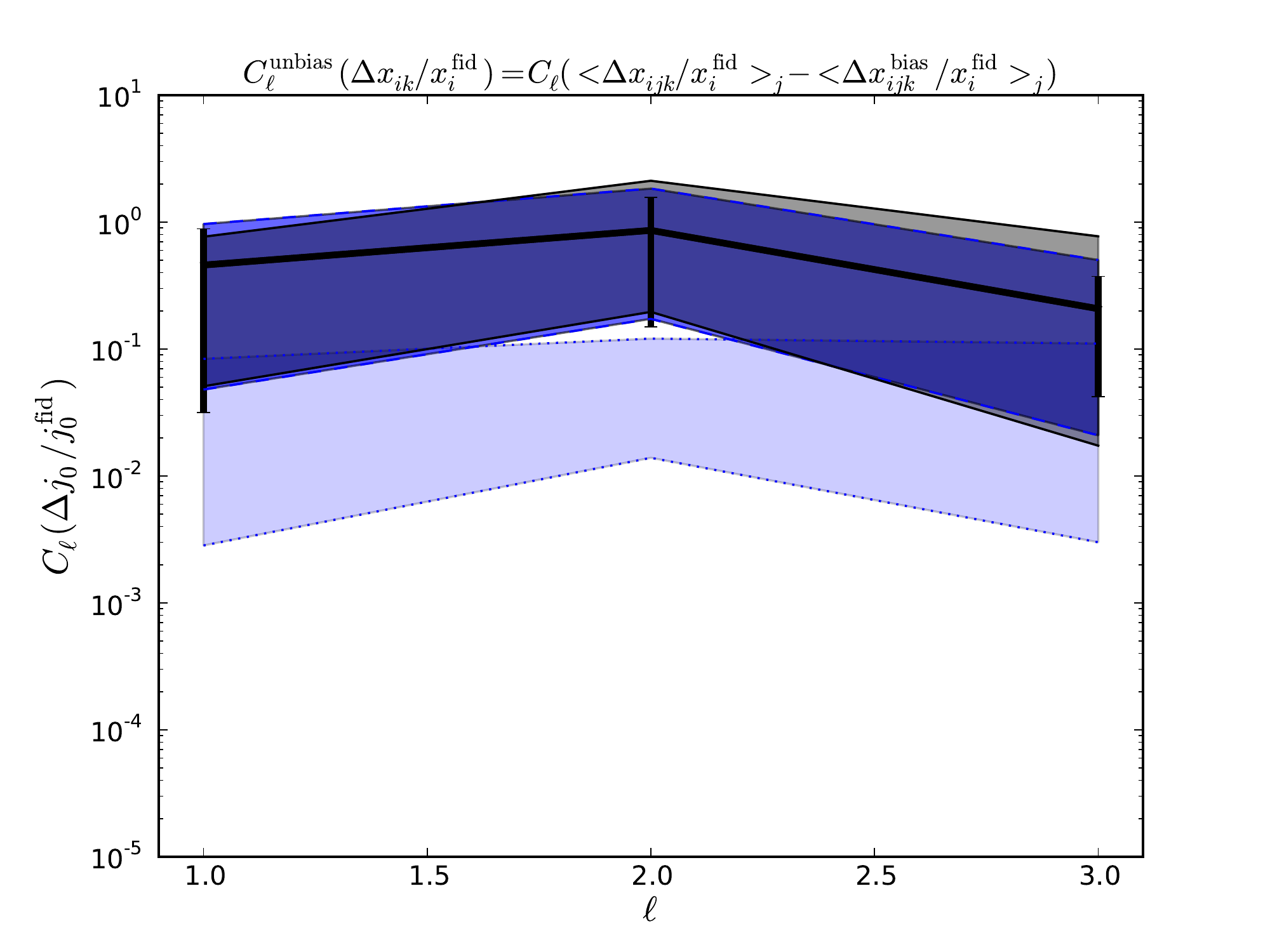}
}
\centerline{
\includegraphics[width=9cm]%{cosmic_var_cl_ldelta_mean_unbias_h0_omegasig0_h0sig1_nchain30_nchainbias100_nsim10000_npix12_var_measure_2.pdf}
{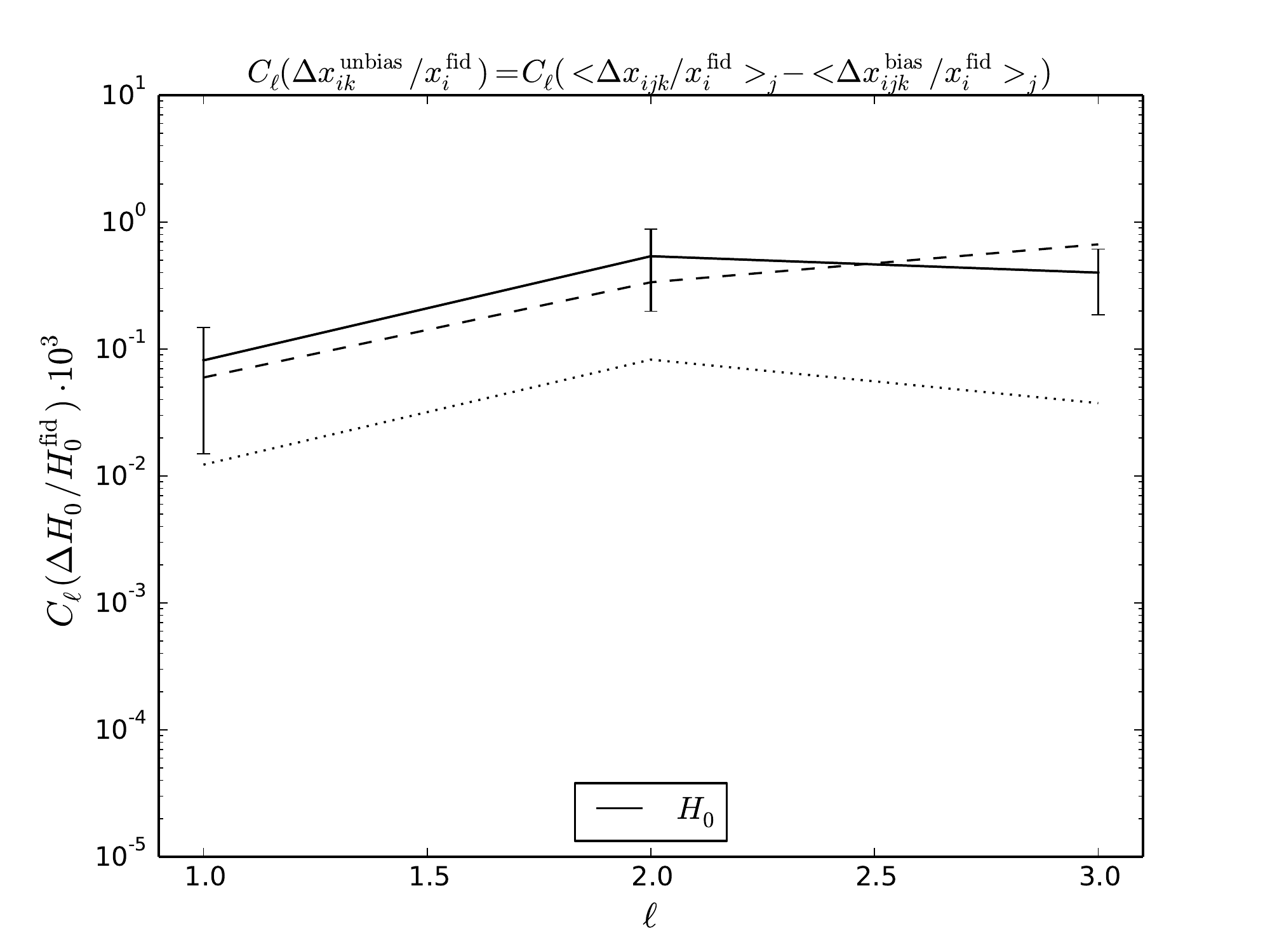}
\includegraphics[width=9cm]%{shift_pix_cosmic_var_fix_k_cl_ldelta_mean_unbias_h0_q0j0sig0_h0sig1_nshift9_nchain300_nchainbias300_nsim10000_npix12_var_measure.pdf}
{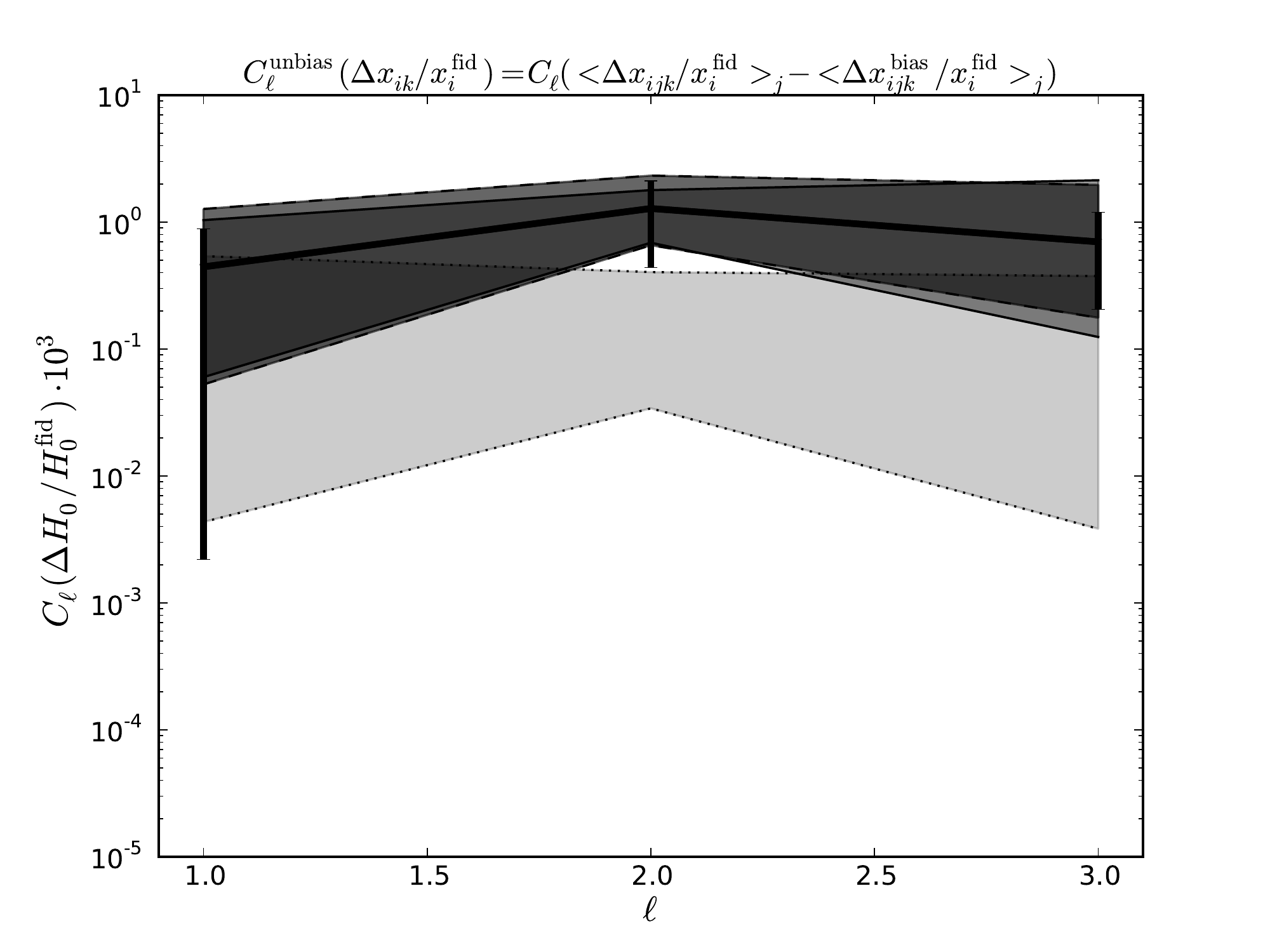}
}
\caption{%\baselineskip=0.4cm{
{\bf Power spectrum of the parameters estimated from the JLA type Ia SN sample (method B).} The dashed lines are the power spectra of the average ``Cosmic variance'' estimation, the dotted lines are the power spectra of the average ``Shuffle SNe'' estimation and the solid lines are the unbiased power spectra defined as the power spectra of the average unbiased maps. Left panels: Power spectra from the $\{H_0,\Omega_{M},\Omega_{\Lambda}\}$ estimation. Right panels: Power spectra from the $\{H_0,q_0,j_0\}$ estimation.%}
}
\label{fig:cl_b_shift}
%\end{figure}
\end{figure*}

\subsection{Power spectrum of unbiased map}

In the second method, we define the unbiased power spectrum as the power spectrum of the mean unbiased maps.
We compute the unbiased power spectrum by computing the power spectrum $ C_{\ell,ip}^{\text{unbias}}$ of the unbiased mean maps $\delta \bar x_{ikp}^{\text{unbias}}=(\bar x_{ikp}-\bar x_{ikp}^{\text{bias}})/x^{\text{fid}},$ with total variance
\ba
\text{Var}[C_{\ell,ip}^{\text{unbias}}]=\text{Var}[ C_{\ell,ip}^{\text{unbias}}]^{\text{estimator}}
.\ea
We then average over the different $p$ pixelations, defining
\ba
C_{\ell,i}^{\text{unbias}}=\big<C_{\ell,ip}^{\text{unbias}}\big>_p,
\ea
with variance
\ba
\text{Var}[C_{\ell,i}^{\text{unbias}}]=\left[\sum_{p=1}^n {1\over \text{Var}[C_{\ell,ip}^{\text{unbias}}]}\right]^{-1}.
\ea
In the right panels of  Fig.~\ref{fig:cl_b_shift}, we plot the resulting unbiased power spectra, for the different pixelations %(as a grey shaded region) 
and the average over the $p$ pixelations. %(as a solid line). 
The variability in $C_{\ell,ip}$ (the power spectrum of the mean map $\delta \bar x_{ikp}=\big<\delta x_{ijkp}\big>_j=\{\delta \bar H_0,\delta \bar q_0, \delta \bar j_0\}_{kp,\text{cosmic\_var}}$ averaged over the $j$ realizations of the ``Cosmic variance'' local estimation) and $C_{\ell,ip}^{\text{bias}}$ (the power spectrum of the mean map $\delta \bar x_{ikp}^{\text{bias}}=\big<\delta x_{ijkp}^{\text{bias}}\big>_j=\{\delta \bar H_0,\delta \bar q_0, \delta \bar j_0\}_{kp,\text{shuffle\_sn}}$ averaged over the $j$ realizations of the ``Shuffle SNe" local estimation) from the different pixelations is indicated as colour--shaded regions bordered by dashed and dotted lines respectively and included for reference.
For all parameters, the power spectrum has a maximum at $\ell=2.$ %which is approximately 
%$\{2.9,2.2,1.9\}$ %[CHECK] %nshift=9,nchain=300,nchain_bias=300
%times the $\ell=1$ mode and 
%$\{1.8,2.0,4.1\}$ %[CHECK] %nshift=9,nchain=300,nchain_bias=300
%times the $\ell=3$ mode in the $\{H_0,q_0,j_0\}$ power spectra respectively.
Although this method has smaller errors than the previous method, the results are still compatible with a flat spectrum.

%[Comparative analysis]

For comparison, in the left panels of Fig.~\ref{fig:cl_b_shift}, using the same line types, we plot the power spectra for the parameters $\{H_0,q_0=\Omega_{M}/2-\Omega_{\Lambda},j_0=\Omega_{M}+\Omega_{\Lambda}\}$ from the $\{H_0,\Omega_{M},\Omega_{\Lambda}\}$ estimation in \citet{carvalho_2015}.
In this estimation, both $H_0$ and $j_0$ have a maximum at $\ell=2,$ whereas $q_0$ decreases always with the multipole.
%The quadrupole is approximately 
%$\{6.6,0.4,141.2\}$ times the dipole 
%and $\{1.3,6.9,123.4\}$ times the octupole 
%in the $\{H_0,q_0=\Omega_{M}/2-\Omega_{\Lambda},j_0=\Omega_{M}+\Omega_{\Lambda}\}$ power spectra respectively. This amounts to a variation in the quadrupole--to--dipole ratio by a factor of 
%$\{0.4,4.9,0.01\}$ %[CHECK] %nshift=9,nchain=300,nchain_bias=300
%and a variation in the quadrupole--to--octupole ratio by a factor of 
%$\{1.4,0.3,0.03\}$ %[CHECK] %nshift=9,nchain=300,nchain_bias=300
Hence, between the $\{H_0,\Omega_{M},\Omega_{\Lambda}\}$ and the $\{H_0,q_0,j_0\}$ estimation, we observe the creation of a peak at $\ell=2$ for $q_0$ and the smoothing of the peak for $H_0$ and $j_0.$ 
We also observe that the power spectra in the $\{H_0,q_0,j_0\}$ estimation are up to
$1\times10^3$ 
times the power spectra in the $\{H_0,\Omega_{M},\Omega_{\Lambda}\}$ estimation, hence yielding an increase of the amplitude. 

The two methods return qualitatively equivalent results; quantitatively, however, the second method might be more efficient at removing the noise bias.

\section{Average values}
\label{sec:average}

In order to compute the average values of the parameters from the corresponding mean maps $\bar x_{ik},$ we average the map of each parameter over the pixel subsamples. 
When averaging over homogeneous pixels, angular fluctuations in the expansion factor (and consequently in $H_{0}$) induce a backreaction term in the average deceleration parameter in the form of an extra positive acceleration \citep{rasanen_2006}. By the same reasoning, angular fluctuations in the expansion factor (and consequently in $H_{0}$ and $q_{0}$) will also induce a backreaction term in the average jerk parameter.
In \citet{carvalho_2015} we derived the analytical extra positive acceleration for a toy model of an arbitrary number of disjoint, homogeneous regions and computed the overall deceleration parameter assuming a) no backreaction and b) backreaction for the measured angular distribution of $H_{0}.$ 
Here, for the same toy model, we derive the corresponding extra terms for the time variation in the acceleration and compute the overall jerk parameter assuming a) no backreaction and b) backreaction for the measured angular distribution of $H_{0}$ and $q_{0}.$ 

In the absence of backreaction, the averaging consists in taking the mean weighted by the variance's inverse $w_k=1/\text{Var}[\bar x_{ik}]$ of parameter $x_{i}$ in each pixel $k,$ %$\bar x_i=\big< \bar x_{ik}\big>_{k},$ 
\ba
\bar x_i=\big< \bar x_{ik}\big>_{k}={\sum_{k}^{N_{\text{pixel}}}w_k~\bar x_{ik} \over {\sum_{k}^{N_{\text{pixel}}}w_k}}.
\ea
For the estimated parameters, we find
$\{\overline H_{0},\overline q_0,\overline j_0\}
=\{70.99,-0.497,0.538\}\pm \{0.54,0.126,0.214\}.$ 
After the noise bias removal, we find 
$\{\overline H_{0},\overline q_0,\overline j_0\}_{\text{unbias}}
=\{71.32,-0.570,0.540\}\pm \{0.60,0.162,0.333\}.$ 
These values are consistent with the fiducial values. After the noise bias removal, the pixel average values 
become closer to the corresponding values estimated using the complete sample.

In the presence of backreaction, the averaging consists in taking the mean weighted by the three--volume $V_k$ of each pixel $k,$ %$\bar x_i=\big< \bar x_{ik}\big>_{V_k},$ 
\ba
\bar x_i=\big< \bar x_{ik}\big>_{V_k}={\sum_{k}^{N_{\text{pixel}}}V_k~\bar x_{ik} \over {\sum_{k}^{N_{\text{pixel}}}V_k}}.
\ea
%Since all pixels have the same surface area and hence the angular directions expand the same way today, we can assume that the radial direction also expands the same way today, which implies that the 3--volume is the same today for all pixels. This amounts to identifying a volume as a pixel. 
Identifying a volume as a pixel and
defining $v_{k}=V_{k}/\sum_{k}V_{k},$ then for $N_{\text{pixel}}$ disjoint regions, the average of the Hubble parameter is given by
\ba
%\left< {\dot a \over a}\right>_{V_k}=\sum_{k}^{N_{\text{pixel}}}v_{k} {\dot a \over a}.
\left< H_0\right>_{V_k}=\sum_{k}^{N_{\text{pixel}}}v_{k} H_{0,k}.
\label{eqn:a_dot_Vk}
\ea

\subsection{Average value of the deceleration parameter}

In order to derive the volume average of the acceleration, we take the time derivative of Eq.~(\ref{eqn:a_dot_Vk}) and find that
\ba
\left<{\ddot a \over a}\right>_{V_k}=
\sum_{k}^{N_{\text{pixel}}}v_{k} {\ddot a \over a} 
%+2\sum_{k}^{N_{{\text pixel}}}v_{k} (1-v_{k})H_{0,k}^2
%-2\sum_{k}^{N_{{\text pixel}}}\sum_{l\not=k}^{N_{{\text pixel}}}v_{k}v_{l}H_{0,k}H_{0,l}
+2\sum_{k}^{N_{\text{pixel}}}\sum_{l>k}^{N_{\text{pixel}}}v_{k}v_{l}\left(H_{0,k}-H_{0,l}\right)^2.
%\equiv \left< {\ddot a \over a}\right>_{V_k}^{\text{lin}}+\left< {\ddot a \over a}\right>_{V_k}^{\text{quad}}.
\label{eqn:a_dotdot_Vk}
\ea
Equation (\ref{eqn:a_dotdot_Vk}) decomposes into a linear term in the pixel average of $(\ddot a/a)$ and a quadratic term in differences of $H_{0}$ between pairs of pixels. The quadratic (backreaction) term generates an acceleration due %not to regions speeding up locally, but instead 
to the slower regions becoming less represented in the average. In the absence of the quadratic term, the volume average reduces to the pixel average above. Then the volume average of $q_{0}$ becomes
\ba
\left< q_{0}\right>_{V_k}&=&\sum_{k}^{N_{\text{pixel}}}v_{k}q_{0,k}\cr
&-&{2\over \left(\sum_{k}^{N_{\text{pixel}}} v_{k}H_{0,k}\right)^2}
\sum_{k}^{N_{\text{pixel}}}\sum_{l>k}^{N_{\text{pixel}}}v_{k}v_{l}\left(H_{0,k}-H_{0,l}\right)^2.
\label{eqn:q0_Vk}
\ea
Using the fluctuations in $H_{0}$ (measured in the ``Cosmic variance'' local estimation) we find 
$\bar q_{0}=-0.498\pm0.126,$ 
and after the noise bias removal %(measured in the ``Shuffle SNe'' local parameter estimation), 
we find 
$\bar q_{0,\text{unbias}}=-0.570\pm0.162$  
(see Table~\ref{table:param_local}).  

The quadratic term is of order 
$7\times 10^{-4}$ 
times the linear term; the corresponding ratio measured in the $\{H_{0},\Omega_{M}, \Omega_{\Lambda}\}$ estimation was of order $10^{-3}$ \citep{carvalho_2015}. The difference due to the backreaction is below the standard deviation, hence unobservable.
It follows that, for the angular fluctuations in $H_{0}$ measured with this SN sample, the contribution of the quadratic term in Eq.~(\ref{eqn:q0_Vk}) is insignificant, which renders the volume averaging equivalent to the pixel averaging. 
These results confirm that, in the context of this toy model of an inhomogeneous space--time and for a kinetic parametrisation, backreaction is not a viable dynamical mechanism to emulate cosmic acceleration.

\subsection{Average value of the jerk parameter}

In order to derive the volume average of the time variation in the acceleration, we take the time derivative of Eq.~(\ref{eqn:a_dotdot_Vk}) and find that  \ba
\left<{\dddot a \over a}\right>_{V_k}&=&
\sum_{k}^{N_{\text{pixel}}}v_{k} {\dddot a_{k} \over a_{k}}\cr 
%+2\sum_{k}^{N_{{\text pixel}}}v_{k} (1-v_{k})H_{0,k}^2
%-2\sum_{k}^{N_{{\text pixel}}}\sum_{l\not=k}^{N_{{\text pixel}}}v_{k}v_{l}H_{0,k}H_{0,l}
&+&2\sum_{k}^{N_{\text{pixel}}}\sum_{l>k}^{N_{\text{pixel}}}v_{k}v_{l}\left(H_{0,k}-H_{0,l}\right)^2
\sum_{m}^{N_{\text{pixel}}}v_{m} H_{0,m}\cr
%&+&2\sum_{k}^{N_{{\text pixel}}}v_{k}(-q_{0,k}H_{0,k}^3)
%-2\sum_{k}^{N_{{\text pixel}}}v_{k}(-q_{0,k}H_{0,k}^2)\sum_{l}^{N_{{\text pixel}}}v_{l}H_{0,l}
&+&2\sum_{k}^{N_{\text{pixel}}}\sum_{l>k}^{N_{\text{pixel}}}v_{k}v_{l}
\left[\left(-q_{0,k}H_{0,k}^2\right)-\left(-q_{0,l}H_{0,l}^2\right)\right]\cr
&&\times\left(H_{0,k}-H_{0,l}\right).
\label{eqn:a_dotdotdot_Vk}
\ea
Equation (\ref{eqn:a_dotdotdot_Vk}) decomposes into a linear term in the pixel average of $(\dddot a/a),$ a quadratic term in differences of $H_{0}$ between pairs of pixels %(similar to the quadratic term in Eq.~(\ref{eqn:a_dotdot_Vk})) 
and a linear term in differences of $(-q_0H_{0}^2)$ between pairs of pixels. The quadratic term in differences of $H_{0}$ generates a contribution that is always positive similar to the quadratic term in Eq.~(\ref{eqn:a_dotdot_Vk}). Conversely, the linear term in differences of $(-q_0H_{0}^2)$ generates a contribution that can have either sign; in particular, it will be positive in the pair of pixels where $(-q_0)$ and $H_0$ vary in the same direction (i.e. both quantities increase or decrease between pixels) and it will be negative in the pair of pixels where $(-q_0)$ and $H_0$ vary in opposite directions (i.e. one quantity increases while the other decreases). Since Eq.~(\ref{eqn:a_dotdotdot_Vk}) measures the angular average of the time variation in the acceleration, the difference (backreaction) terms generate an extra jerk that is due to slower regions and/or more slowly varying regions becoming less represented in the average. Similarly to Eq.~(\ref{eqn:a_dotdot_Vk}), in the absence of the difference terms, the volume average reduces to the pixel average above. Then the volume average of $j_{0}$ becomes
\ba
\left<j_{0}\right>_{V_k}&=&
\sum_{k}^{N_{\text{pixel}}}v_{k} {j_{0,k}} \cr
&+&{2\over \left(\sum_{k}^{N_{\text{pixel}}} v_{k}H_{0,k}\right)^2}
\sum_{k}^{N_{\text{pixel}}}\sum_{l>k}^{N_{\text{pixel}}}v_{k}v_{l}\left(H_{0,k}-H_{0,l}\right)^2\cr
&-&{2\over \left(\sum_{k}^{N_{\text{pixel}}} v_{k}H_{0,k}\right)^3}\cr
&&\times\sum_{k}^{N_{\text{pixel}}}\sum_{l>k}^{N_{\text{pixel}}}v_{k}v_{l}
\left(q_{0,k}H_{0,k}^2-q_{0,l}H_{0,l}^2\right)\left(H_{0,k}-H_{0,l}\right).
\label{eqn:j0_Vk}
\ea
Using the fluctuations in $\{H_{0},q_{0}\}$ (measured in the  ``Cosmic variance'' local parameter estimation), we find 
$\bar j_{0}=0.546\pm 0.218$  %[CHECK] %nshift=9,nchain=300,nchain_bias=300
and after the noise bias removal, %(measured in the ``Shuffle SNe'' local parameter estimation), 
we find 
$\bar j_{0,\text{unbias}}=0.548\pm 0.345$  %[CHECK] %nshift=9,nchain=300,nchain_bias=300
(see Table~\ref{table:param_local}). 

The quadratic term in differences of $H_{0}$ is of order 
$1\times 10^{-3}$  %[CHECK] %nshift=9,nchain=300,nchain_bias=300
times the linear term, whereas the linear term in differences of $(-q_0H_{0}^2)$ is of order 
$1\times 10^{-2}$  %[CHECK] %nshift=9,nchain=300,nchain_bias=300
times the linear term; the corresponding ratios in the $\{H_{0},\Omega_{M}, \Omega_{\Lambda}\}$ estimation in \citet{carvalho_2015} were of order $6\times 10^{-4}$  and $2\times 10^{-3}.$ 
%for $\Omega_{\kappa}$ free. %, and $10^{-7}$  and $2\times 10^{-6}$ for $\Omega_{\kappa}=0.$ 
%Consequently, the total difference (backreaction contribution) is of order 10 times that measured in the $\{H_{0},\Omega_{M}, \Omega_{\Lambda}\}$ estimation. 
Since $j_{0}$ is more poorly constrained than $q_{0}$ or $H_{0}$ in the global estimation, the total difference due to the backreaction is still below the standard deviation, hence unobservable. It follows that, for the angular fluctuations in $\{H_{0},q_{0}\}$ measured with this SN sample, the contribution of the backreaction terms in Eq.~(\ref{eqn:j0_Vk}) is insignificant, which renders the pixel averaging equivalent to the pixel averaging. These results imply that 
an inhomogeneous $j_0,$ such that at different pixels the acceleration changed at different times, 
cannot be distinguished from a globally homogeneous $j_0,$ such that the acceleration changed everywhere at the same time.

For comparison, we present the pixel average of $\{q_0=\Omega_{M}/2-\Omega_{\Lambda},j_0=\Omega_{M}+\Omega_{\Lambda}\}$ from the $\{H_{0},\Omega_{M}, \Omega_{\Lambda}\}$ estimation in \citet{carvalho_2015}, both in the absence and in the presence of backreaction, which we include in Table~\ref{table:param_local}. 

\begin{table*}%[t]
\caption{%\baselineskip=0.4cm{
{\ Values for the parameters estimated from the JLA type Ia SN sample.} %Column 1: The parameters estimated either directly or indirectly from the fit. Column 2: The values estimated from the complete SN sample. Columns 3--4: The values estimated from the subsampling of SNe into pixels of equal surface area, before and after the noise bias subtraction. Column 5: The averaging method.%}
}
\label{table:param_local}
\begin{tabular}{cc||cccc}
\hline\hline
&Parameter 
& {Complete sample} 
& \multicolumn{2}{c} {Subsample into pixels} & Averaging\\ 
&&  & Biased  & Unbiased & \\ \hline%\hline
\multirow{5}{*}{$\kappa=0\!:$} & $H_{0}$ 
& $71.06\pm 0.46$ 
& $70.99\pm 0.54$ & $71.32\pm 0.59$ & $\left<H_{0}\right>_k$\\
& \multirow{2}{*}{$q_0$}
& \multirow{2}{*}{$-0.540\pm 0.094$} 
& $-0.497\pm 0.126$ & $-0.570\pm 0.162$ & $\left<q_{0}\right>_k$\\
&&&$-0.498\pm 0.126$ & $-0.570\pm 0.162$ &$\left<q_{0}\right>_{V_k}$\\
&\multirow{2}{*}{$j_0$}
& \multirow{2}{*}{$0.533\pm 0.503$} 
& $0.538\pm 0.214$ & $0.540\pm 0.333$ & $\left<j_{0}\right>_k$\\
&&&$0.546\pm 0.218$ & $0.548\pm 0.345$ &$\left<j_{0}\right>_{V_k}$\\
\hline
\multirow{5}{*}{$\Omega_{\kappa}=1-\Omega_{M}-\Omega_{\Lambda}$:} & $H_0$ 
%\multirow{5}{*}{$\Omega_{\kappa}$ free:} & $H_0$ 
& $71.17\pm 0.44$
& $71.06\pm 0.87$ & $71.48\pm 0.94$ & $\left<H_{0}\right>_k$\\
& \multirow{2}{*}{$\Omega_{M}/2-\Omega_{\Lambda}$} 
& \multirow{2}{*}{$-0.586\pm 0.123$} 
&  $-0.451\pm 0.159$ & $-0.527\pm 0.172$ & $\left<q_{0}\right>_k$\\
&&&$-0.452\pm 0.159$ & $-0.528\pm 0.173$& $\left<q_{0}\right>_{V_k}$\\
& \multirow{2}{*}{$\Omega_{M}+\Omega_{\Lambda}$} 
& \multirow{2}{*}{$0.971\pm 0.139$} 
&  $0.922\pm 0.131$ & $0.996\pm 0.135$ & $\left<j_{0}\right>_k$\\
&&&$0.926\pm 0.134$ & $0.998\pm 0.138$ & $\left<j_{0}\right>_{V_k}$\\
\hline
\multirow{5}{*}{$\Omega_{\kappa}=0\!:$} & $H_0$ 
& $71.21\pm 0.33$
& $70.93\pm 0.64$ & $71.22\pm 0.90$ & $\left<H_{0}\right>_k$\\
& \multirow{2}{*}{$\Omega_{M}/2-\Omega_{\Lambda}$} 
& \multirow{2}{*}{$-0.599\pm 0.020$} 
&  $-0.586\pm 0.061$ & $-0.600\pm 0.083$ & $\left<q_{0}\right>_k$\\
&&&$-0.586\pm 0.061$ & $-0.600\pm 0.083$& $\left<q_{0}\right>_{V_k}$\\
& \multirow{2}{*}{$\Omega_{M}+\Omega_{\Lambda}$} 
& \multirow{2}{*}{$1.000\pm 0.026$} 
&  $1.000\pm 0.011$ & $1.000\pm 0.011$ & $\left<j_{0}\right>_k$\\
&&&$1.002\pm 0.010$ & $1.000\pm 0.0134$& $\left<j_{0}\right>_{V_k}$\\
\hline
\end{tabular}
\tablefoot{Column 1: The parameters estimated either directly or indirectly from the fit. Column 2: The values estimated from the complete SN sample. 
Columns 3--4: The values estimated from the subsampling of SNe into pixels of equal surface area, before and after the noise bias subtraction. Column 5: The averaging method.
}
%\caption{\label{table:ena} {\bf The data. Latest}}
\end{table*}

\section{Conclusions}
\label{sec:concl}

%[UPDATE] 
In this paper we used SN data to fit a kinematic parametrisation of the luminosity distance expressed in terms of time derivatives of the scale factor. This parametrisation records the cosmic expansion without regard to its cause, thus being independent of the cosmological equation of state and consequently adequate to test interpretations of the cosmic acceleration alternative to the cosmological constant. We followed the parameter estimation, first presented in \citet{carvalho_2015}, to fit the parameters $\{H_0,q_0,j_0\}$ by performing both a global and a local parameter estimation. 
From the global parameter estimation, using the complete SN sample, we obtained the fiducial values adopted in this manuscript. From the local parameter estimation, dividing the SNe into subsamples over a pixelated map, we obtained maps of the estimated parameters with the same pixelation as the SN subsamples. 
We then proceeded to the analysis of this paper's results as well as to a comparative analysis with the results from \citet{carvalho_2015} estimated by fitting $\{H_0,\Omega_{M},\Omega_{\Lambda}\}$ instead.

The measurements of $\{H_0,q_{0},j_{0}\}$ are consistent with %, and have errors of the same order as, 
the measurements of $\{H_{0},q_{0}=\Omega_{M}/2-\Omega_{\Lambda},j_{0}=\Omega_{M}+\Omega_{\Lambda}\}$ obtained in \citet{carvalho_2015}. However, whereas the error of  $q_{0}$ is of the same order in both parametrisations (about a $5\sigma$ measurement), the error of $j_{0}$ is significantly larger in the kinematic parametrisation (from a $7\sigma$ to a $1\sigma$ measurement). This is a consequence of the kinematic parametrisation in part minimising the physical assumptions that enter in the model and in part truncating the Taylor expansion of the luminosity distance. 

We measured fluctuations about the average values of order 
0.1--5\% %[CHECK] %nshift=9,nchain=300,nchain_bias=300
for $H_0,$ 
1--150\% %[CHECK] %nshift=9,nchain=300,nchain_bias=300
for $q_0$ and  
1--124\% %[CHECK] %nshift=9,nchain=300,nchain_bias=300
for $j_0.$ Comparing with the fluctuations measured in \citet{carvalho_2015}, we observe an increase by a factor of 
%order 22--124.
two orders of magnitude.
We also computed the power spectrum of the corresponding maps of the parameters up to $\ell=3,$ as determined by the pixel size, finding that all power spectra have a maximum at $\ell=2,$ regardless of the method used to subtract the noise bias. This observation can be partially ascribed to the absence of objects towards the galactic plane. %Comparing with the power spectra measured in Ref.~\citep{carvalho_2015}, we observe a smoothing of the peak at $\ell=2$ and an increase by up to a factor of $10^2$ in the amplitude of the angular correlation.

Finally, for an analytical toy model of an inhomogeneous ensemble of homogenous pixels, we measured the backreaction term in $q_0$ due to the fluctuations of $H_0$ to be of order 
$5\times 10^{-4}$ %[CHECK] %nshift=9,nchain=300,nchain_bias=300
the corresponding pixel average in the absence of backreaction, hence of smaller order than that measured in \citet{carvalho_2015}. We also derived the backreaction term in $j_0$ due to the fluctuations of $\{H_0,q_0\}$ and measured it to be of order 
$1\times 10^{-2}$ %[CHECK] %nshift=9,nchain=300,nchain_bias=300
the corresponding pixel average in the absence of backreaction, hence 
about 50 times  
that measured in \citet{carvalho_2015}. 
However, both backreaction effects are below the corresponding standard deviation and hence rendered unobservable. It follows that backreaction generates insignificant extra acceleration to emulate cosmic acceleration, and insignificant extra jerk to emulate different late--time cosmic histories.
 
An interesting idea would be to extend this type of analysis to further inhomogeneity studies by combining SN data with other cosmic tracers, such as high--redshift galaxies \citep{terlevich_2015} and galaxy clusters \citep{bengaly_2015b,bolejko_2015}.
\\

\noindent{\bf Acknowledgements} The authors thank A. Afonso and N. Nunes for useful discussions. CSC is funded by Funda\c{c}\~ao para a Ci\^encia e a Tecnologia (FCT), Grant no. SFRH/BPD/65993/2009. 

%%%%%%%%%%%%%%%%%%%%%%%%%%%%%%%%%%%%%%%%%%%%%%%%%%

%%%%%%%%%%%%%%%%%%%% REFERENCES %%%%%%%%%%%%%%%%%%

\end{document}